\documentclass{osa-article}

\journal{oe}

 \usepackage{parskip}

\usepackage[utf8]{inputenc} 
\usepackage[T1]{fontenc}
\usepackage{graphicx}
\usepackage{grffile}
\usepackage{longtable}
\usepackage{wrapfig}
\usepackage{rotating}
\usepackage{grffile} % remove spaces in filenames for figures
\usepackage[normalem]{ulem}
\usepackage{amsmath}
\usepackage{textcomp}
\usepackage[title]{appendix}
\usepackage{capt-of}
\usepackage{hyperref}
\usepackage{subcaption}
\usepackage{marginnote}
\usepackage{setspace}
\usepackage{float}
\usepackage{cases}

\usepackage{physics}

\newcommand{\neff}{n_{\text{eff}}}
\newcommand{\heff}{h_{\text{eff}}}

\newcommand{\lcwl}{\lambda_{\text{cwl}}}

\newcommand{\conj}[1]{\overline{#1}}
\usepackage{epstopdf}

\usepackage{tikz}
\usepackage[mode=buildnew]{standalone}
\usepackage{fp,tikz,tikz-3dplot}
\usepackage{tkz-euclide}
\usetikzlibrary{decorations.pathreplacing,decorations.markings}
\usetikzlibrary{quotes,angles}
\usetikzlibrary{patterns,3d,calc}

\newcommand{\E}{{\mathcal{E}}}

\newcommand{\HH}{{\mathcal{H}}}

\newcommand{\Tray}{T_{\text{tiny}}^{\text{ray}}}

\newcommand{\Twp}{T_{\text{tiny}}^{\text{wave}}}

\newcommand{\apptinywp}{\ref{app:tinytransmittance}}
\newcommand{\appcountingM}{\ref{app:countingM}}
\newcommand{\appfwhm}{\ref{app:fwhm}}

\newcommand{\appanalytical}{\ref{app:analytical}}
\newcommand{\apppeaktransmittance}{\ref{app:peak}}
\newcommand{\apptradeoff}{\ref{app:tradeoff}}
\newcommand{\appfdfd}{\ref{sec:FDFD}}
\newcommand{\appexperiment}{\ref{app:experiment}}

\begin{document}

\title{Pixel-integrated thin-film filter simulation and scaling trade-offs} 

\author{Thomas Goossens\authormark{1,2,*}}
\address{\authormark{1}Stanford University\\
\authormark{2} This work was partially performed at imec (Belgium) and Institut Fresnel (France)}
\email{\authormark{*}Email:  contact@thomasgoossens.be} %% email address is required

\begin{abstract*}
Thin-film optical filters can nowadays be integrated onto pixels of commercial image sensors used for spectral imaging. A drawback of having more filters on an image sensor is a loss in spatial resolution which could be regained by using smaller pixels. However, this work shows that miniaturization causes a reduction in transmittance and requires a trade-off between pixel size and filter bandwidth.
This is caused by diffraction and the reduced number of reflections in the filter.
Analytical wave and ray optics models are developed and used for fast transmittance prediction and this is validated using a commercial spectral sensor and numerical simulations.
The identified trade-off is an intrinsic limitation of pixel-integrated thin-film filter technology and the developed models will be useful tools for exploring new hardware and computational solutions.
\end{abstract*}

\section{Introduction}

% Spectral imaging and applications
Spectroscopes obtain a spectrum for a single point and use this to identify and analyze different materials \cite{skoog2017principles}.
While useful, such devices remain  unpractical for analyzing and comparing the spectrum across large areas.
In contrast, spectral imaging techniques combine imaging and spectroscopy such that the spectrum at different positions can be compared simultaneously. The applications range from, but are not limited to, monitoring plant health for precision agriculture \cite{Khan2018}, endoscopy \cite{Yoon2019}, recycling \cite{Serranti2015}, and art preservation \cite{Khan2018}.

\begin{figure}[htpb!]
\centering 
\includegraphics[width=0.8\linewidth]{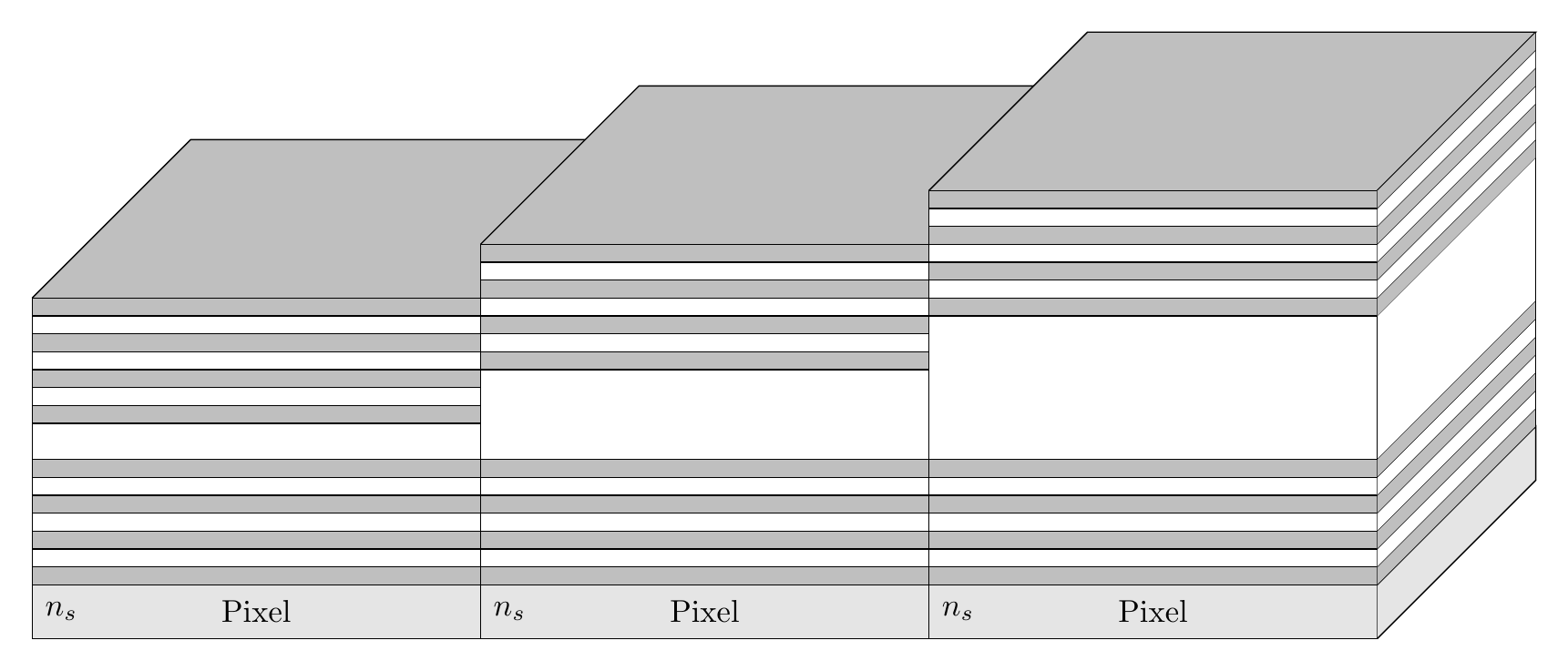}
\caption{\label{fig:array} Array of pixel-integrated thin-film Fabry-Pérot filters made from two dielectric materials. By changing the cavity thickness, the transmitted wavelengths can be chosen. }
\end{figure}

% Trend in cameras
Over the decades, spectral imaging devices have become smaller and lighter to facilitate new applications \cite{Hagen2013}. 
A major advance in miniaturization was enabled by the deposition of thin-film spectral filters on 5.5 µm wide pixels of commercial CMOS image sensors (Fig.~\ref{fig:array}) \cite{Geelen2014,Geelen2014a,Geelen2015}.
The monolithic integration of filters on the pixels makes the cameras more mechanically robust, compact, and lightweight.
Another recent advancement is the development of short-wave infrared spectral cameras using an InGaAs sensor with 15 µm wide pixels \cite{Gonzalez2018}.
While the scope of this work is limited to integrated thin-film filters, many alternative technologies are actively being investigated \cite{Williams2019,Shaukat2020,Kusar2020}. 

\begin{figure}[H]
  \centering
  \begin{subfigure}[t]{0.35\linewidth}
    \includegraphics[width=0.99\linewidth]{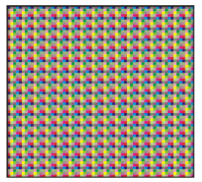}
    \caption{\label{fig:mosaic} Extended Bayer (mosaic) pattern.}
  \end{subfigure}
\hfill
  \begin{subfigure}[t]{0.35\linewidth}	
    \includegraphics[width=0.9\linewidth]{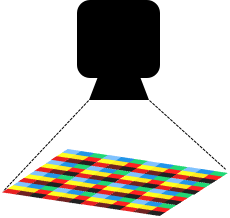}
    \caption{\label{fig:mosaicoperation} The mosaic camera acquires a spectral image in one shot.}
  \end{subfigure}
  \caption{\label{fig:mosaicpattern} The extended Bayer (mosaic) pattern and its mode of operation. Images at the courtesy of imec.}
\end{figure}

In applications that require high frame rates or video capture, the filters can be deposited in an extended Bayer (mosaic) pattern. Like the Bayer pattern \cite{Bayer1976}, a cell of \(N\) by \(N\) filters is periodically repeated across the sensor (Fig.~\ref{fig:mosaicpattern}) \cite{Geelen2014,Geelen2015}.
A major drawback of this approach is the loss in spatial resolution by a factor of $N$ in each dimension. While the 5.5 µm resolution might be sufficient for some applications, there is a high demand for snapshot spectral cameras with higher spatial resolution, rather than more spectral bands. One hardware solution would be to deposit filters on smaller pixels.

To design pixel-integrated thin-film filters, good simulation tools are required. However, most commercial thin-film software tools are based on the transfer-matrix method which assumes an infinite filter size \cite{macleod2017,optilayer,macleod1997essential,tfcalc}; An assumption which, as will be shown, produces wrong transmittance predictions at present pixel scales.

As such, this work makes two contributions to the field of pixel-integrated filter design.
First, the development of wave- and ray-optics models that include pixel size and that can predict filter transmittance and its angular dependency.
Second, the derivation of a fundamental scaling trade-off between pixel size and filter bandwidth to avoid a given reduction in filter transmittance.

These findings are particularly relevant in the context of spectral cameras with pixel-integrated filters.
Firstly, because the scaling trade-off constrains the possible filter designs and what optics can be used.
And secondly, because the software tools developed in this work could become useful tools to efficiently explore different filter designs. 
 
The manuscript is organized as follows.
An overview of related work is presented in Section~\ref{sec:related}.
A wave-optics model is derived in Section~\ref{sec:theory}.
An equivalent monolayer model is introduced in Section~\ref{sec:equivalentmodel} and used to derive
a ray-optics interpretation in Section~\ref{sec:raymodel}.
This leads to the formulation of a trade-off law in Section~\ref{sec:tradeoff}.
The results are validated numerically and experimentally in Section~\ref{sec:validation}.
Finally, the analysis is concluded with a discussion on trade-off mitigation strategies and future directions.

\section{Tiny thin-film filters and related work}
\label{sec:related}
In the thin-film filter literature, the term \textit{thin} is loosely defined as a film which is thin enough for interference effects to be observed \cite[p38]{macleod2017}. Similarly, the term \textit{tiny} thin-film filter will be used to indicate that a filter is spatially narrow enough to produce deviations from classical thin-film theory which supposes an infinite filter width. The term narrow was refrained from to avoid confusion with the term \textit{narrowband} filter, which is already very common.

No discussion on this topic occurs in recent reference works on thin-film filters \cite{macleod2017,amra2021electromagnetic} and commercial software tools like OptiLayer \cite{optilayer}, TFCalc \cite{tfcalc}, or Essential Macleod \cite{macleod1997essential} do not model the effect of finite width.
There is remotely related work on modeling the effect of gaps in a mirror of a Fabry-Pérot etalon using a ray model. Except for suggesting the usefulness of ray models, it remains different from studying the finite width of multi-layered filters \cite{GEIGER1949,VanderSluis1956}. 
Other relevant work includes the study on pixel vignetting by Catrysse et al. \cite{Catrysse2002}. In
their appendix it is shown how a finite pixel size causes a loss of rays that lowers the transmittance for increasing angles. This pixel-vignetting model, however, only predicts a scaling of pixel response due to a loss in flux. In contrast, this manuscript deals with changes in spectrum due to changes in the interference pattern in filters deposited onto a pixel.

The wave-optics model that will be introduced is mathematically related to illuminating an infinite film with a light beam of narrow width. This small-spot illumination is briefly discussed in Macleod's handbook using the angular spectrum \cite[p. 614]{macleod2017}.
The formalism is mathematically similar but there are some important differences with this work.
First, in this work the finite size of the filter is incorporated by integrating the transmitted flux only across the finite pixel area. Second, the formalism is given a new interpretation using a ray model.

\section{Wave-optics approach }
\label{sec:theory}

This section introduces a wave-optics model used for tiny-filter simulation. The detailed derivivations are documented in Section~\apptinywp~of Supplementary Document 1.

\begin{figure}[htpb!]
\centering
\includegraphics[width=0.8\linewidth]{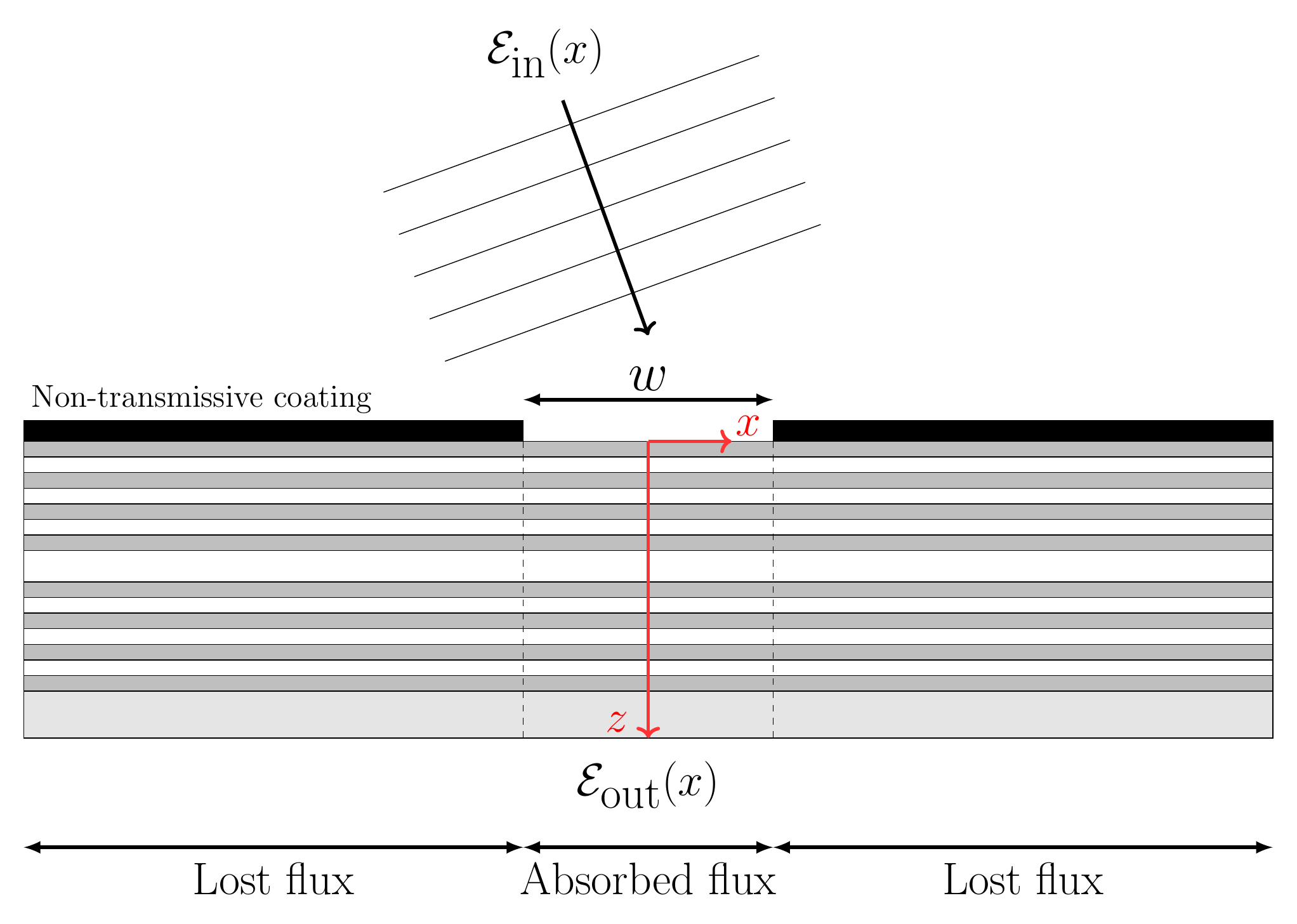}
\caption{\label{fig:infapprox} Plane wave incident on an infinite film coated with an aperture. Only the flux across the pixel array is considered to contribute.}
\end{figure}

To approximate the transmittance of a tiny filter, it is assumed that the finite size can be modeled by having an aperture on an infinitely wide thin-film filter stack (Fig.~\ref{fig:infapprox}).
In the context of imaging we are then interested in the total power transmitted to the pixel (substrate). All flux arriving outside of the pixel area is considered lost with respect to its contribution to the original interference pattern. This lost flux might contribute to cross-talk but this is outside the scope of this work.

The effect of finite filter size is that an incident plane wave will be spatially constrained such that
\begin{equation}
 \label{eq:Ein}
\mathcal{E}_\text{in}(x) = \text{rect}\left(\dfrac{x}{w}\right)\underbrace{e^{2\pi i x\sin\theta/\lambda}}_{\text{Plane wave}}\,,
\end{equation}
where \(\text{rect}(\frac{x}{w}) = 1\, \text{for}\, |x/w|<1/2\).
The factor \(2\pi i x\sin\theta/\lambda\) accounts for the phase difference that arises because the wave front arriving at position \(x\) has to travel an additional
distance \(x\sin\theta\) compared to the wave front arriving at \(x=0\).

To analyze how this field is transmitted, we can represent it using its angular spectrum \cite{Goodman2005}, also referred to as a wave packet decomposition \cite{amra2021electromagnetic}. This means the incident field can be decomposed in plane waves by means of a spatial Fourier transform such that the plane wave amplitudes are
\begin{subequations}
\label{eq:angularspectrum}  
\begin{align}
A_\text{in}(\nu) =& \int_{-\infty}^{+\infty} \mathcal{E}_\text{in}(x) e^{-i(2\pi\nu x)} \textup{d}x,\\
  =& w\,\text{sinc}\left( \pi w \left(\nu - \dfrac{\sin\theta}{\lambda}\right)\right),
\end{align}
\end{subequations}
with $\text{sinc}(x)=\sin(x)/x$. The spatial frequencies $\nu$ correspond to the different incidence angles of the plane waves. For example, for an incident plane wave $e^{2\pi ix\sin\theta/\lambda}$, we have $\nu = \frac{\sin\theta}{\lambda}$. 

Now the transmission coefficient $t$ for plane wave illumination can be used to calculate how each plane wave in the decomposition is transmitted  \cite[p.126]{amra2021electromagnetic}.
The total transmitted field is then reconstructed as the integral
over the transmitted plane waves such that 
\begin{equation}
\label{eq:transmittedfield}
\E_t(x,\lambda) = \int_{-\infty}^{+\infty}\underbrace{t(\nu,\lambda)A_\text{in}(\nu,\lambda)}_{\displaystyle A_t = tA_{\text{in}}}e^{i(2\pi\nu x)} \textup{d}\nu,
\end{equation}
which corresponds to taking the inverse Fourier transform.

Using the wave-packet approach, the tiny-filter transmittance $\Twp$ is calculated as the total transmitted
power \(\Phi_\text{t}\) [W] divided by the total incident power \(\Phi_\text{in}\) [W] so that
\begin{equation}
  \label{eq:Twp}
  \Twp(\lambda;\theta) =  \dfrac{\Phi_\text{t}}{\Phi_\text{in}}.
\end{equation}
 
A full derivation is presented in Supplementary Document 1 (Section~\apptinywp) and leads to the following expressions which can be efficiently implemented 
\begin{align}
  \Phi_{\text{in}} =& \dfrac{\eta_{\text{in}}}
                      {2} \displaystyle\int_{-\frac{w}{2}}^{\frac{w}{2}}   \conj{\E_\text{in}}(x)  \E_\text{in}(x) \dd{x} = \dfrac{\eta_{\text{in}}w}{2},  \\
  \Phi_\text{t}                  =&  \dfrac{1}{2}  \int_{\nu'}  \eta_t(\nu')A_\text{t}(\nu') (\conj{A_{\text{t}}} * K)(\nu') \textup{d}\nu'       \label{eq:phi-t},
\end{align}
with $K(\nu')= w \text{sinc}(\pi \nu'w)$,  $*$ the convolution operator, and $\eta$ [$AV^{-1}$] the (polarization dependent) characteristic admittance of each layer \cite[p.112]{amra2021electromagnetic}.
When necessary, the convolution can be computed efficiently using a Fast-Fourier Transform.

\section{Equivalent tiny monolayer model}
\label{sec:equivalentmodel}

In \eqref{eq:transmittedfield} and \eqref{eq:phi-t}, only the transmission coefficient $t$ in $A_{t}=tA_{\text{in}}$ contains filter-specific information.  Therefore, if $t$ can be approximated in a certain wavelength and angular range of interest, the model gives identical results. This important property will be essential for developing a ray-optics interpretation and deriving a fundamental scaling trade-off.

To find an equivalent transmission coefficient we employ the well-established fact that all-dielectric Fabry-Pérot filters can be approximated by an equivalent monolayer model in the neighborhood of the main peak \cite{Pidgeon1964,macleod2017}. However, its application to the study of tiny filters is a novel contribution of this work.

The transmission coefficient for an infinitely wide cavity with refractive index $\neff$ and mirrors of reflectance $R$ mirrors is derived in
\cite[p.362]{born1999} so that
\begin{equation}
  t =  \dfrac{t_s (1-R)}{1-R e^{i2\delta}}, 
\end{equation}
which is used to set $A_t  =t A_{\text{in}}$ in \eqref{eq:phi-t}. Here, $\delta = \frac{2\pi \neff \heff \cos \theta_n }{\lambda}$ is the phase thickness, $\theta_n$ the angle of refraction, and $t_s$ the transmission coefficient between the incident medium and the substrate.

So essentially, only two model parameters are required: the mirror reflectance $R$ and the effective refractive index $\neff$ of the cavity. The central wavelength $\lcwl$ of the filter at normal incidence can be fixed by choosing a half-wave cavity thickness $\heff = \frac{\lcwl}{2\neff}$. 

The reflectance $R$ of the mirrors is directly related to the normalized Full Width at Half Maximum (FWHM) $\Lambda_\infty$ of the \textit{infinitely wide} Fabry-Pérot filter as
\begin{equation}
  \label{eq:alpha}
  R \sim 1-\pi\Lambda_\infty,\quad \Lambda_\infty = \dfrac{\text{FWHM}}{\lcwl}\rightarrow 0,
\end{equation}
as derived in Section~\appfwhm~of Supplementary Document 1.

The effective refractive index $\neff$ fully determines how much the filter shifts towards shorter wavelengths for increasing incidence angles.
When the filter design is known, an analytical expression for the effective index can be used \cite[p.279]{macleod2017}. For a low-index cavity we have
\begin{equation}
\label{eq:neff}
\neff = n_l\left(1-\dfrac{n_l}{n_h} + \dfrac{n_l^2}{n_h^2}\right)^{-\frac{1}{2}},
\end{equation}
where $n_h$ is a high refractive index $n_l$ a low refractive index.
Alternatively, when the filter design is not known, the effective index can also be estimated empirically by measuring the central wavelength at multiple angles \cite{Goossens2019d}.

In the next section, the equivalent tiny monolayer model is used to develop a ray-optics interpretation and an analytical expression for the transmittance.

\section{Ray optics interpretation of tiny filters}
\label{sec:raymodel}
It is already known that ray models give correct results for infinitely wide filters illuminated by plane waves \cite{macleod2017}. In contrast, using ray analysis to study tiny filters is proposed as a new approach which offers valuable insights into the effect of finite filter size.
We limit ourselves to the analysis of a single-cavity filter which corresponds to the equivalent monolayer model introduced in the previous section.

\subsection{Truncated interference}

\begin{figure}[htpb!]
 \centering
\includegraphics[width=0.9\linewidth]{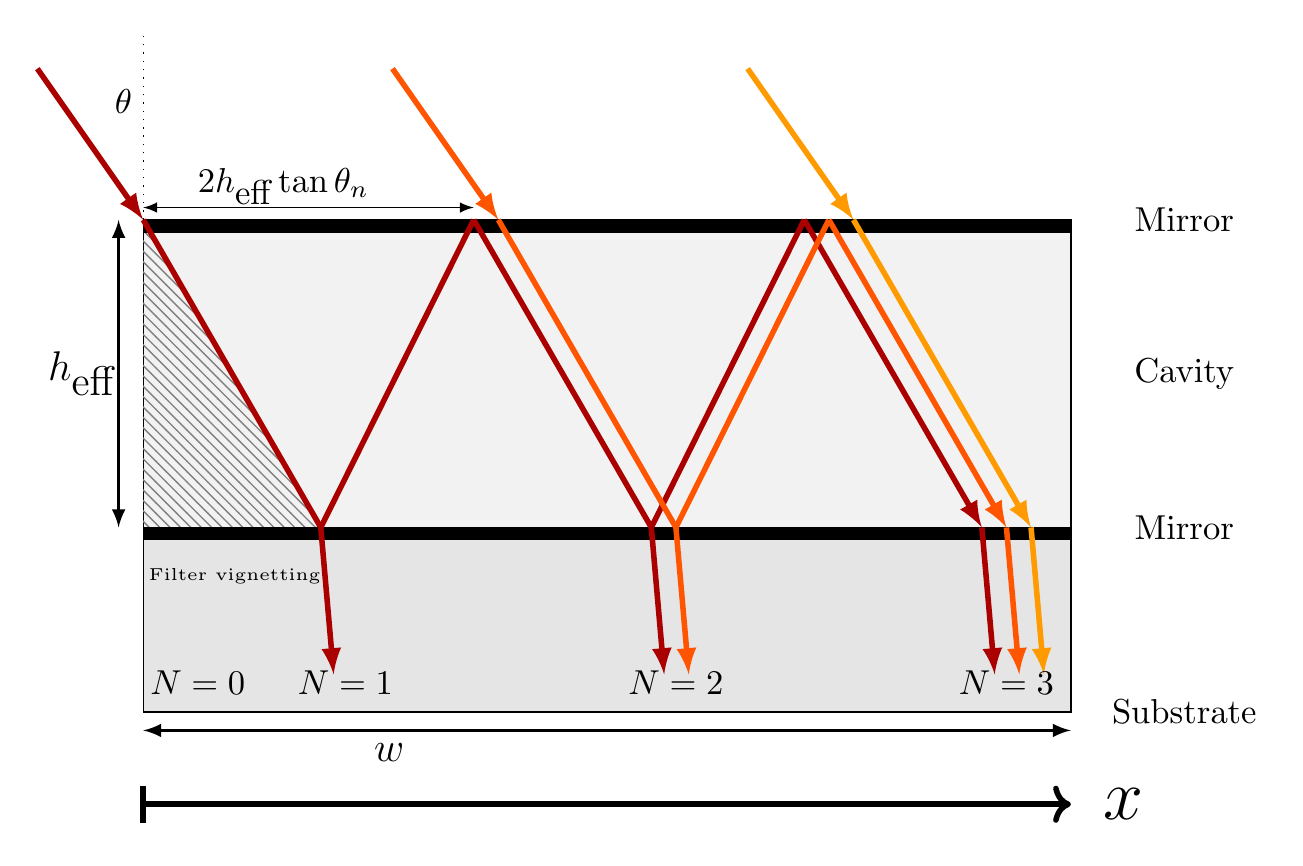}
\caption{\label{fig:tinymonolayer} In a tiny Fabry-Pérot filter, only a finite number $N(x)$ of rays can interfere at position $x$ on the substrate. The overlapping rays are drawn next to each other for illustration purposes only.}
\end{figure}

The correct operation of a thin-film interference filter depends on light beams reflecting many times and interfering with each other. When the filter is infinitely wide, there are an infinite number of rays that interfere.
In contrast, when the filters are tiny, the supply of rays at oblique incidence is limited but increases with distance from the wall (Fig.~\ref{fig:tinymonolayer}). This means that the number of interfering rays varies across the substrate.

For example, consider the ray labeled as $N=1$ in Fig.~\ref{fig:tinymonolayer}. Starting from the left wall, this is the first ray that can be transmitted through the filter. This means to the left of this ray, no light arrives and this can be referred to as filter vignetting, by analogy to pixel vignetting \cite{Catrysse2002}.
In contrast, the region between between $N=1$ and $N=2$ only involves direct transmission without interference. In the subsequent regions, gradually more rays start interfering.

This interpretation is formalized by calculating the resulting field amplitude at each position on the substrate. For classical thin-film theory, this leads to a converging infinite series \cite[p.361]{born1999}. However, when there are only \(N(x)\) outcoming rays at position $x$, the series has to be truncated such that
\begin{subequations}
\begin{eqnarray}
\label{eq:seriestrans}
  \E_t(x;\lambda) &=&  t_s t^2 \sum_{n=1}^{N(x)} R^{(n-1)}  e^{i2\delta(n-1)}\\
                  &=& t_s t^2\cdot  \left( \dfrac{R^{N}e^{i2N\delta}-1}{Re^{i2\delta}-1}\right),
\end{eqnarray}
\end{subequations}
where $t$ is the transmission coefficient of the mirror, $r$ the reflection coefficient, $R=\conj{r}r$, and $t_s$ the transmission coefficient of the air-substrate interface.
The factor 
\begin{equation}
  \delta = \frac{2\pi \neff \heff \cos(\theta_n)}{\lambda},
\end{equation}
is defined as the \emph{phase thickness} of the film. It accounts for the relative differences in optical path length that each ray travels \cite{born1999}.

The physical interpretation of the series is that an incident ray of unit amplitude is first partially transmitted (\(t\)), then
traverses through the layer to the lower interface where it is partially reflected
(\(r\)). The ray then traverses the layer to the top interface where it is, again, partially
reflected (\(r\)). This process continues until the ray reaches the position $x$ where it is partially transmitted through the mirror and to the substrate ($tt_s$). The transmission coefficients are outside the summation because each ray that enters the substrate has entered and exited the cavity only once.

The position-dependent number of interfering rays is equal to
\begin{equation}
  \label{eq:M}
N(x) = \left\lfloor\dfrac{1}{2} + \dfrac{x}{2\heff\tan\theta_n}\right\rfloor,
\end{equation}
as derived in Supplemental Document 1 (Section~\appcountingM). Here, $\lfloor\cdot\rfloor$ is the floor operator which rounds number downwards.
This number becomes smaller for increasing incidence angles because the
horizontal displacement $\heff\tan\theta_n$ for each traversal becomes larger. Hence, rays reach the boundary with fewer reflections in the cavity.

\subsection{Transmittance of a tiny filter}
\label{sec:tiny-effectivemono}

% For an infinitely wide filter illuminated by a plane wave, the
% irradiance is position independent. By letting \(M\rightarrow\infty\) in \eqrefn{eq:geomtrans}, we have
% \begin{eqnarray}
% \label{eq:tinf-exact}
%   T_\infty^{\text{ray}}(\lambda) &=& \dfrac{\eta_2}{\eta_0} \dfrac{(t_{01}t_{12})^2}{(1-2r_{12}r_{10}\cos(2\delta)+(r_{12}r_{10})^2)},
% \end{eqnarray}
% which is known as one of Airy's formulae \cite{born1999}.
% This infinite-film transmittance $T_\infty(\lambda)$ will be used for comparison with the effective transmittance
% of a tiny filter.
For an infinitely wide filter, there is a position-independent transmitted irradiance. 
For a tiny filter, in contrast, the irradiance \(I(x;\lambda)\) [W\,m\textsuperscript{-2}] is position \textit{dependent}. As in \eqref{eq:Twp}, the transmittance of a tiny filter is then defined as the total transmitted
power divided by the total incident power so that
\begin{equation}
\label{eq:transintegrated}
\Tray(\lambda) 
 = \dfrac{\displaystyle \int_0^w I_t(x;\lambda) \dd x}{\displaystyle \int_0^w I_\text{in}(x;\lambda) \dd x}.
\end{equation}

An analytical approximation for the filter transmittance is derived (Section~\appanalytical~of Supplemental Document 1) so that
\begin{align}
  \label{eq:analyticalray} 
   \Tray(\lambda) &\approx  \frac{\eta_s}{\eta_{\text{in}}}\cdot T_s (1-R)^2 \cdot \frac{\left[1 + \frac{R^{2M}-1}{\log R^{2M}} - \frac{2(\log(R) (R^M\cos(2M\delta)-1)+ 2\delta R^M\sin(2M\delta))}{4M\delta^2+M\log(R)^2}\right]}{1-2R \cos(2\delta)+R^{2}},
\end{align}
with 
\begin{equation}
  \label{eq:M-ray}
M \approx \dfrac{w\neff}{\lcwl\tan\theta_n},
\end{equation}
being the maximal number of interfering rays in the substrate.

It then follows that the peak transmittance is described by
\begin{equation}
\label{eq:peakvalue}
   \dfrac{T_{\text{peak}}^{\text{ray}}}{T^\infty_{\text{peak}}} =  1+ \dfrac{(1-R^M)(3-R^M)}{\log R^{2M}},
\end{equation}
with $T^\infty_{\text{peak}}= T_{\text{peak}}^{\text{ray}}(M\rightarrow\infty)$.

This equation can be used to quickly estimate the expected drop in peak transmittance for an arbitrary Fabry-Pérot filter. The normalization by the $T^\infty_{\text{peak}}$ enables comparison between filters with different peak transmittances, e.g. caused by using different substrate materials.

\section{Filter size and bandwidth trade-off}
\label{sec:tradeoff}
The equivalent monolayer model introduced in Section~\ref{sec:equivalentmodel} suggests that any Fabry-Pérot filter design that maps onto the same equivalent model parameters will have the same reduction in transmittance. This, of course, only in the wavelength and angular range where the equivalent model is a valid approximation.

An important corollary is the existence of a fundamental trade-off between filter size and filter bandwidth. That is, that for a given requirement in peak transmittance or bandwidth, one cannot freely choose pixel size and filter bandwidth. Or alternatively, that for further miniaturization, larger filter bandwidths might be required.

The trade-off is imposed by two, mostly independent, effects: truncated interference and diffraction which are discussed in the sections that follow.

\subsection{Contribution of truncated interference}
\label{sec:contribtruncate}
Truncated interference was introduced as a ray-optics concept in Section~\ref{sec:raymodel}. It states that for a tiny filter, a converging infinite series representing interfering rays has to be truncated, causing an increase in bandwidth and decrease in peak transmittance.

For narrowband filters ($\Lambda_\infty\rightarrow 0$), the peak transmittance can be approximated by
\begin{equation}
\label{eq:peakvaluehighfinesse}
\dfrac{T_{\text{peak}}^{\text{ray}}}{T^\infty_{\text{peak}}}   \sim 1- \dfrac{(1-e^{-\pi M\Lambda_\infty})(3-e^{-\pi M\Lambda_\infty})}{2\pi M\Lambda_\infty},
\end{equation}
see Section~\apppeaktransmittance~of Supplemental Document 1. 
It follows that the transmittance is fully determined by the factor $M \Lambda_\infty$ which can be rewritten as the following dimensionless trade-off parameter
\begin{equation} 
  \begin{array}{lcl}
  \label{eq:tradeoff}
  \alpha&=&\dfrac{1}{M \cdot \Lambda_\infty} = W^{-1}\cdot\Lambda_\infty^{-1} \cdot \Theta\\\\
        &=&   \underset{\substack{\text{Normalized}\\\text{spatial width}\\\\W}}{\Bigl(\underbrace{\dfrac{w}{\lcwl}}_{}\Bigr)^{-1}} \cdot
  \underset{\substack{\text{Normalized}\\\text{bandwidth}\\\\\Lambda_\infty}}{\Bigl(\underbrace{\dfrac{\text{FWHM}}{\lcwl}}_{}\Bigr)^{-1}}
  \cdot
  \underset{\substack{\text{Normalized}\\\text{incidence angle}\\\\\Theta}}{\Bigl(\underbrace{\dfrac{\tan \theta_n}{\neff}}_{}\Bigr)}
\end{array}.
\end{equation}
The factor $\alpha$ is defined as the inverse of $M\Lambda_\infty$ because this makes it proportional to the incidence angle which facilitates interpretability.
The third factor can be understood as a normalized incidence angle because of the small-angle approximation  $\tan \theta_n \sim \theta/\neff$. This last approximation is not used in software implementations.

For increasing incidence angles $\theta$, the peak transmittance will follow the curve given by \eqref{eq:peakvaluehighfinesse} (Fig.~\ref{fig:peaklaw}). How far is traveled along the curve is fully determined by the system parameters that determine alpha $\alpha$.
In any case, the larger $\alpha$, the lower the transmittance of the filter at oblique incidence due to a corresponding truncation of interference.

% \begin{figure}[H] 
%   \centering
% \includegraphics[width=0.99\linewidth]{/home/thomas/Documents/tinyfilters/research/formulas/peakvalue/fig/peaklaw.png}
%   \caption{\label{fig:peaklaw} Reduction in peak transmittance plotted as a function of two non-dimensional parameters: the normalized FWHM and the normalized angle. After a transition region the filters quickly converge to the corresponding curve predicted by \eqref{eq:peakvalue}}
% \end{figure}

\subsection{Verification with the wave-optics model}

The nondimensionalization suggests that any Fabry-Pérot filter which maps onto the same $\alpha$ value, will have a identical reduction in transmittance and increase in FWHM.
This claim is tested using the wave-optics model, and this for four filters with different material parameters, number of layers, and bandwidth $\Lambda_\infty$ (See Section~\apptradeoff~of Supplementary Document 1).
The peak transmittance is calculated for unpolarized light incident at angles between 0 and 25 degrees and for pixel sizes of $w=5.5$ µm and $w=10$ µm. The peak transmittance and change in FWHM are plotted as a function of $\alpha$, calculated using \eqref{eq:tradeoff} (Figs.~\ref{fig:peaklaw} and \ref{fig:fwhmlaw}).
As expected, not each curve travels equally far horizontally because of the nondimensionalization.

The peak transmittance for both $w=5.5$ µm and $w=10$ µm matches well with the predictions from \eqref{eq:peakvalue}.
There is, however, a width-dependent transition region for small angles. This is due to diffraction and its contribution is explained in the next section.

Similar conclusions appear to hold for the change in FWHM, normalized by the FWHM at normal incidence (Fig.~\ref{fig:fwhmlaw}). For the 5.5 µm pixel, there is a clear deviation from truncated interference but the trend is similar. For the 10 µm pixel, all curves are almost perfectly aligned.

\subsection{Contribution of diffraction}
\label{sec:contribdiffract}

\begin{figure}[htbp!]
  \centering
  \begin{subfigure}[t]{0.4\linewidth}
    \includegraphics[width=0.99\linewidth]{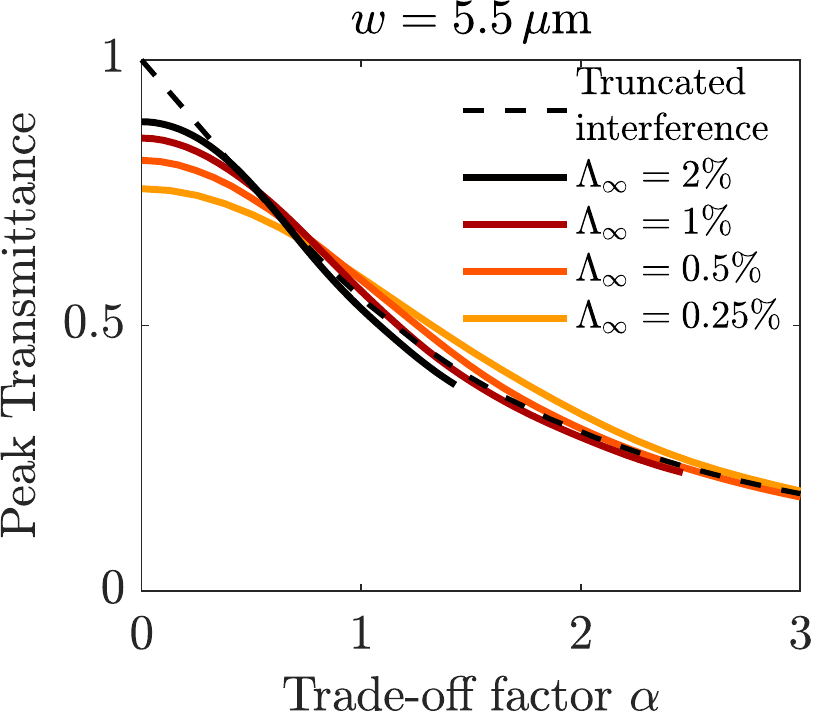}
  \end{subfigure}
  \begin{subfigure}[t]{0.4\linewidth}
    \includegraphics[width=0.99\linewidth]{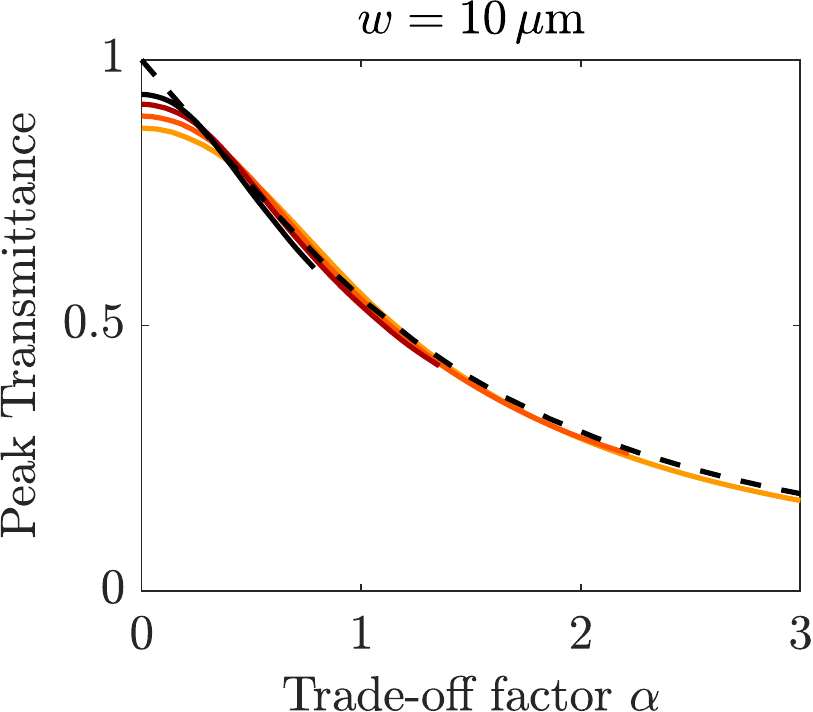}
  \end{subfigure}
  \caption{\label{fig:peaklaw} \textbf{Peak transmittance} plotted as a function of the dimensionless trade-off factor. The truncated interference model works best for the 10 µm pixel. The full lines are wave-optics calculations.}
\end{figure}

\begin{figure}[htpb!]
  \centering
  \begin{subfigure}[t]{0.4\linewidth}
    \includegraphics[width=0.99\linewidth]{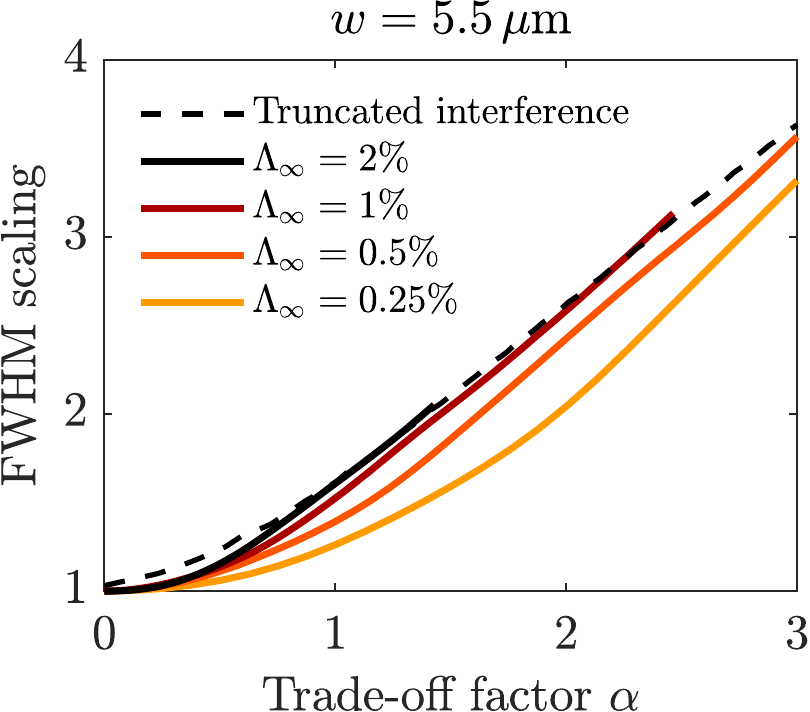}
  \end{subfigure}
  \begin{subfigure}[t]{0.4\linewidth}
    \includegraphics[width=0.99\linewidth]{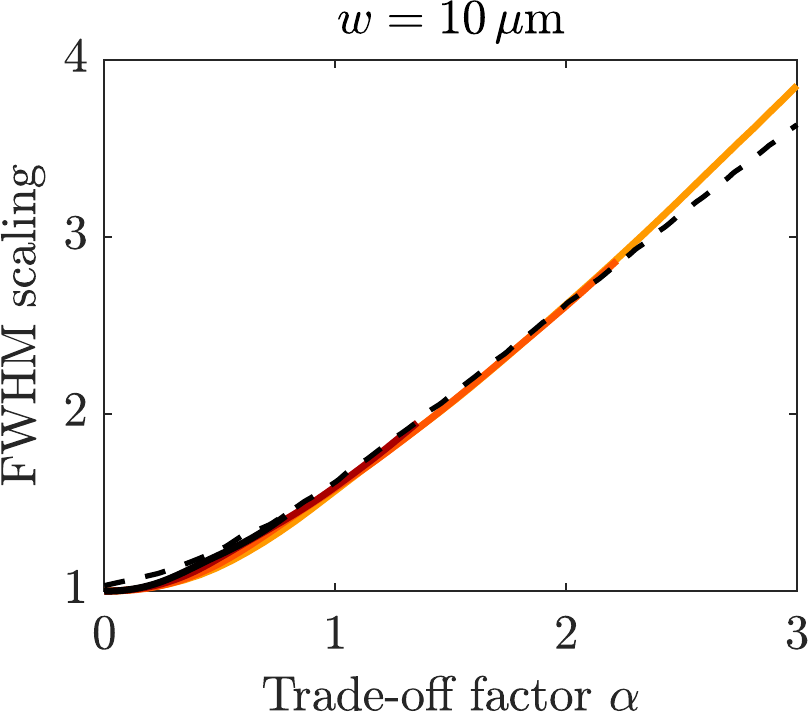}
  \end{subfigure}
  \caption{\label{fig:fwhmlaw} \textbf{Increase in FWHM} plotted as a function of the dimensionless trade-off factor.  The truncated interference model works best for the 10 µm pixel. The full lines are wave-optics calculations.}
\end{figure}

The contribution of diffraction dominates when $\Lambda_\infty$ and $W$ are small (Figs.~\ref{fig:peaklaw} and \ref{fig:fwhmlaw}). As will be numerically validated in the next section, the transmittance-reducing effect of diffraction occurs and is accurately predicted by the wave-optics model.
However, further analysis is required to find simple analytical approximations that can predict this peak drop.

The intuition is that when $W$ is small, the angular spectrum of the diffracted light beam has a wider support (see  \eqref{eq:angularspectrum}). Because the filter transmittance shifts with increasing angle, the effect of diffraction corresponds to taking a weighted average of shifted filter responses. When the bandwidth $\Lambda_\infty$ is small, the averaging effect is stronger.


\section{Validation}
\label{sec:validation}
In this section the wave- and ray-optics models are validated numerically and experimentally.
First, numerically to validate the model at different pixel scales (Section \ref{sec:fdfd}). Second, experimentally using the equivalent monolayer model (Section~\ref{sec:experiments}) and by demonstrating the trade-off law (Section~\ref{sec:tradeoff-experiment}).
 
\subsection{Numerical validation}
\label{sec:fdfd}
To validate the method for multiple pixel sizes, Maxwell's equations are solved using MaxwellFDFD, a finite difference frequency domain (FDFD) Matlab toolbox \cite{maxwellfdfd}. The simulations are performed in two dimensions.

The filters are of a typical all-dielectric Fabry-Pérot filter design found in the literature \cite{macleod2017}:
\begin{equation}
  \label{tf:stacksim}
\text{Air}|H (LH)^b| L_c | (HL)^b H|\text{Silicon},
\end{equation}
where each layer is a quarter-wave plate for a chosen wavelength and $LH$ indicates a layer of low and then high refractive index. The exponent in $(LH)^b$ indicates this pattern is repeated $b$ times.

Three filters are placed adjacently similar to Fig.~\ref{fig:array}.
Only the transmittance of the central filter is calculated to ensure realistic boundary conditions. 
The simulation was run for pixel sizes of 10 µm, 5.5 µm, and 2 µm. More detailed information about the simulations can be found in Section~\appfdfd~of Supplementary Document 1.

For all simulated pixel sizes, the infinite-filter prediction $T_\infty$, fails to explain the change in filter transmittance.
In contrast, for the 10 µm and 5.5 µm wide pixels, the predicted tiny-filter transmittance $\Twp$ corresponds well to the numerical result. 
For the 2 µm pixel, the model predicts most of the variation but there is is a small shift in central wavelength and some non-negligible resonances are visible near the main peak.
Yet, even for the 2 µm pixel, $\Twp$ is much more accurate than the infinite-filter prediction $T_\infty$.

The ray model matches the transmittance well for pixel sizes of 5.5 µm and 10 µm. For the 2 µm pixel, the prediction is bad at normal incidence but good at large angles.
The reason the ray model fails at normal incidence is that the effect of diffraction is not included.

\begin{figure}[hptb!]
  \centering
    \begin{subfigure}[t]{0.9\linewidth}
    \includegraphics[width=0.99\linewidth]{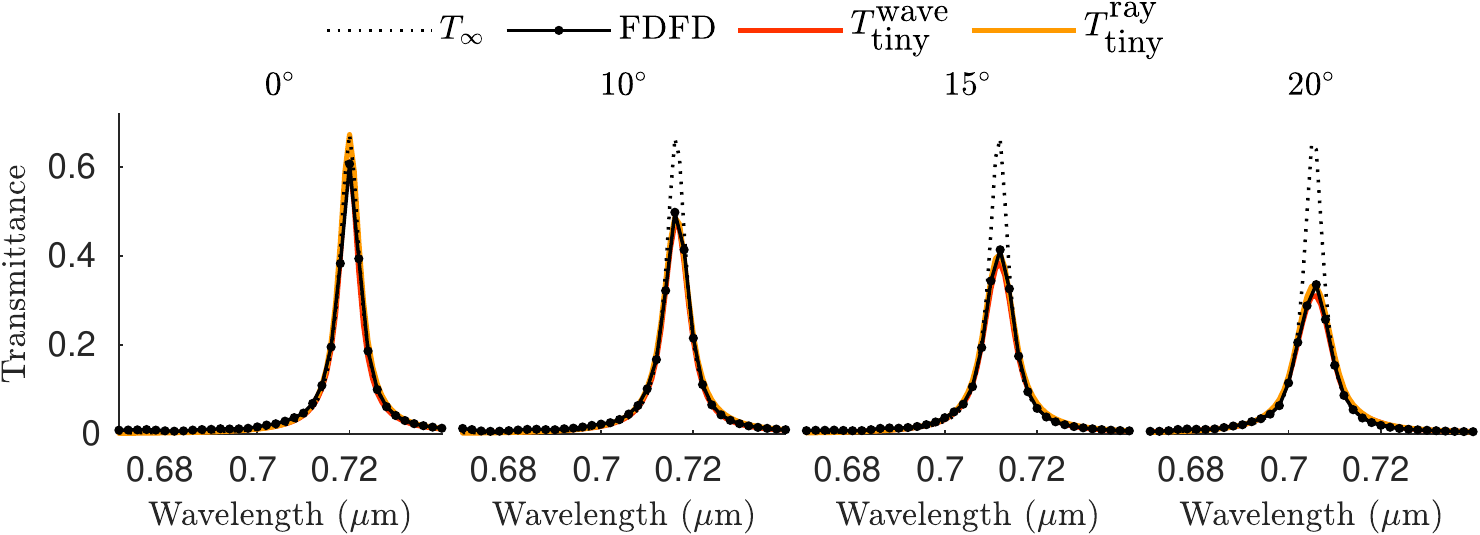}
    \caption{\label{fig:fdfd-10micron} 10 µm pixel}
  \end{subfigure}
  \begin{subfigure}[t]{0.9\linewidth} 
    \includegraphics[width=0.99\linewidth]{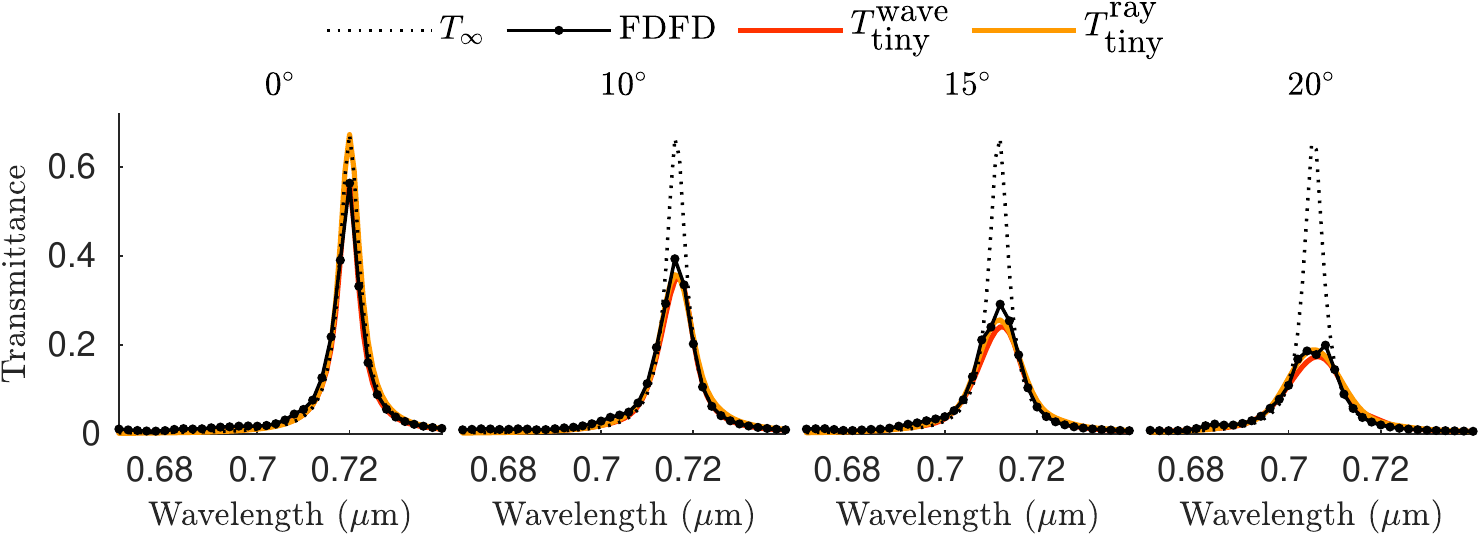}
    \caption{\label{fig:fdfd-5micron} 5.5 µm pixel}
  \end{subfigure}
  \begin{subfigure}[t]{0.9\linewidth}
    \includegraphics[width=0.99\linewidth]{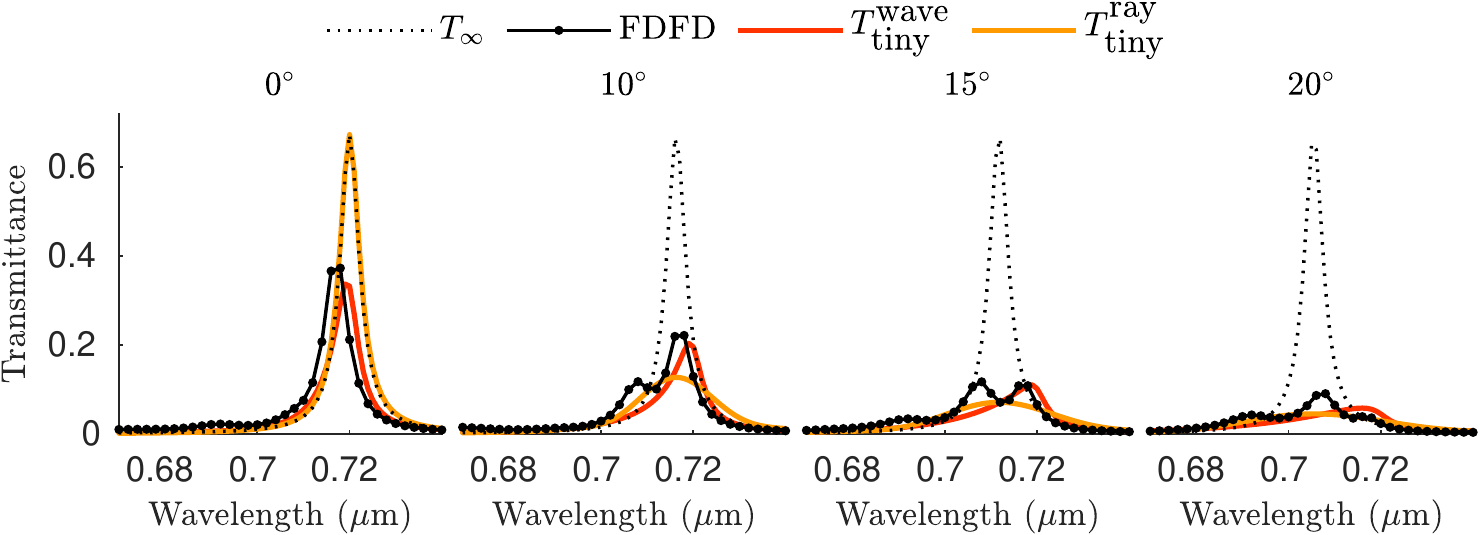}
    \caption{\label{fig:fdfd-2micron} 2 µm pixel}
  \end{subfigure}
  \caption{\label{fig:fdfd} FDFD simulations of the transmittance for different pixel sizes for s-polarized light. The numerical model simulates a 2D version of Fig.~\ref{fig:array} and is compared to predictions of the wave-optics and ray-optics model.}
\end{figure}

\subsection{Experimental validation using the equivalent monolayer model}
\label{sec:experiments}
The models are experimentally validated for a commerical snapshot spectral sensor developed by imec. It consists of  pixel-integrated Fabry-Pérot filters deposited on 5.5 µm wide pixels of a CMV2000 image sensor \cite{Tack2012,CMOSIS}.
Direct validation of the model would require imec's proprietary filter stack, which is not publicly available.
However, using the equivalent monolayer approach we can still make predictions. Detailed information on the experimental setup and data processing is presented in Section~\appexperiment~of Supplemental Document 1.

For imec's sensor, the effective refractive index is taken to be $\neff=1.7$ as reported in previous studies on the angular dependency \cite{Goossens2018}.
The bandwidth $\Lambda_\infty$ is estimated by fitting the wave-optics prediction to the measurement at normal incidence. 
For the first filter located at $\lcwl=767$ nm this gives $\Lambda_\infty=0.91\%$ and for the second filter at $\lcwl=921$ nm that $\Lambda_\infty=1.63\%$.

For the filter at $\lcwl=767$ nm, the peak maximum gradually reduces with incidence angle with a 77\% reduction at $20^\circ$. In addition, the bandwidth (Full-Width at Half Maximum) of the filter more than doubles from $7$ nm to $17$ nm. Recently, similar result were independently observed by Hann et al. \cite{Hahn2020}.

\begin{figure}[htpb!]
  \centering
  \begin{subfigure}[t]{0.9\linewidth} 
    \includegraphics[width=0.99\linewidth]{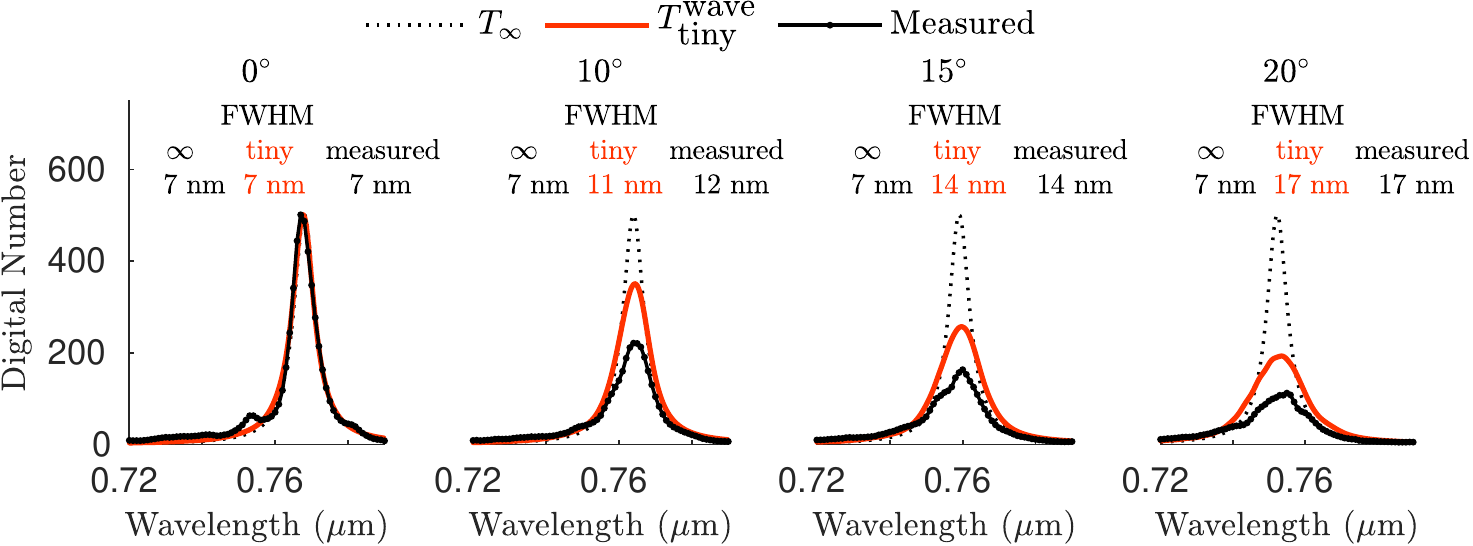}
    \caption{\label{fig:simulations-mosaic} Filter centered at 767 nm. The change in FWHM is predicted well while the peak transmittance is slightly overestimated.}
  \end{subfigure}
  \begin{subfigure}[t]{0.9\linewidth} 
        \includegraphics[width=0.99\linewidth]{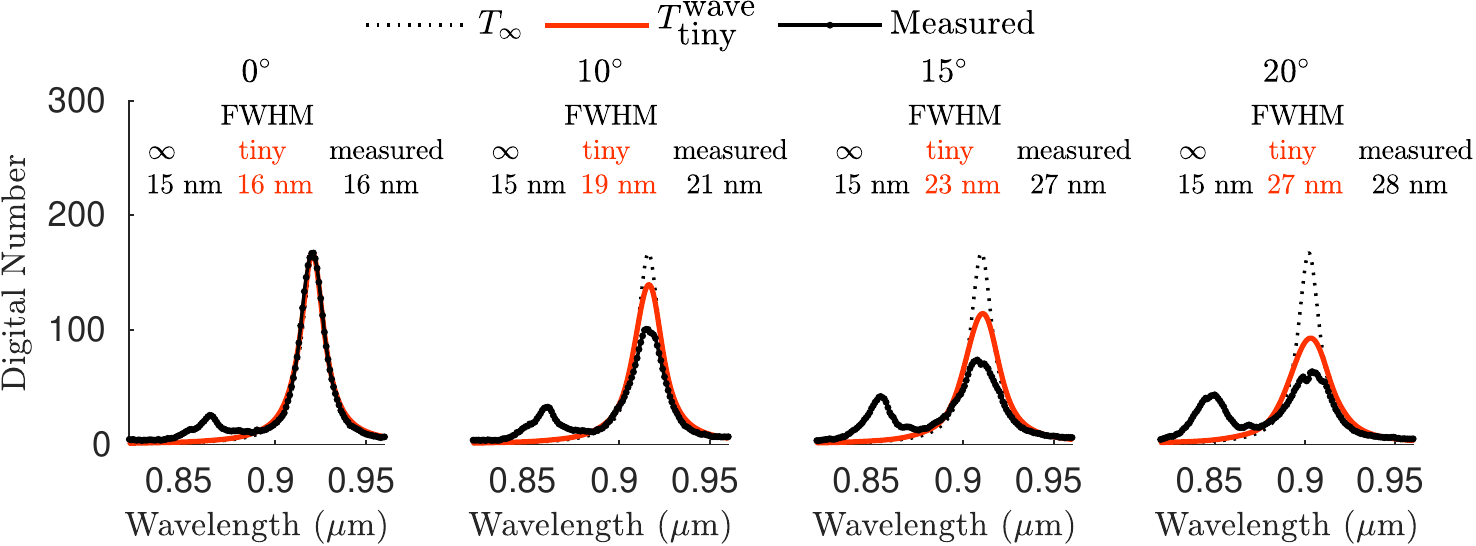}
    \caption{\label{fig:simulations-mosaic2} Filter centered at 921 nm with a higher FWHM. The change in FWHM is predicted well. A second peak, most likely due to cross-talk, is seen around 0.85 µm.}
  \end{subfigure}
  \caption{\label{fig:simulations} Comparison of predicted and measured filter response for two filters on imec's mosaic sensor with two different bandwidth (FWHM). The dashed line is the classical prediction for an infinitely wide filter. In both cases, the tiny filter prediction is much better.}
\end{figure}

For both filters, at all angles, the tiny filter prediction $\Twp$ is much better than the infinite filter prediction $T_\infty$ (Fig.~\ref{fig:simulations}). 
Although the predicted peak is slightly higher than the measurement, the increase in FWHM is predicted accurately. 
This suggests that another effect, like cross-talk, further diminishes the pixel response. In support of this hypothesis, we see in Fig.~\ref{fig:simulations-mosaic2} a second peak at 0.85 µm that increases with the angle and corresponds to the wavelength of the neighboring filter. However, further analysis of cross-talk is outside the scope of this work.

Finally, it is remarkable that such a good fit can be obtained without knowing the actual filter design. 
The impact of this result is reinforced in the next section which validates the trade-off law for imec's sensor.

\subsection{Filter size and bandwidth trade-off law}
\label{sec:tradeoff-experiment}
The trade-off relation is validated for all 25 filters on imec's 5x5 snapshot mosaic sensor used in Section~\ref{sec:experiments}. The filters are distributed across a wavelength range of 665 and 975 nm.

For each filter, at each angle, its trade-off parameter $\alpha$ was calculated using \eqref{eq:tradeoff}.
The bandwidth $\Lambda_\infty$ of each filter was found by fitting the equivalent monolayer wave-optics model to the measurement at normal incidence. 

\begin{figure}[htpb!]
  \centering
    \includegraphics[width=0.99\linewidth]{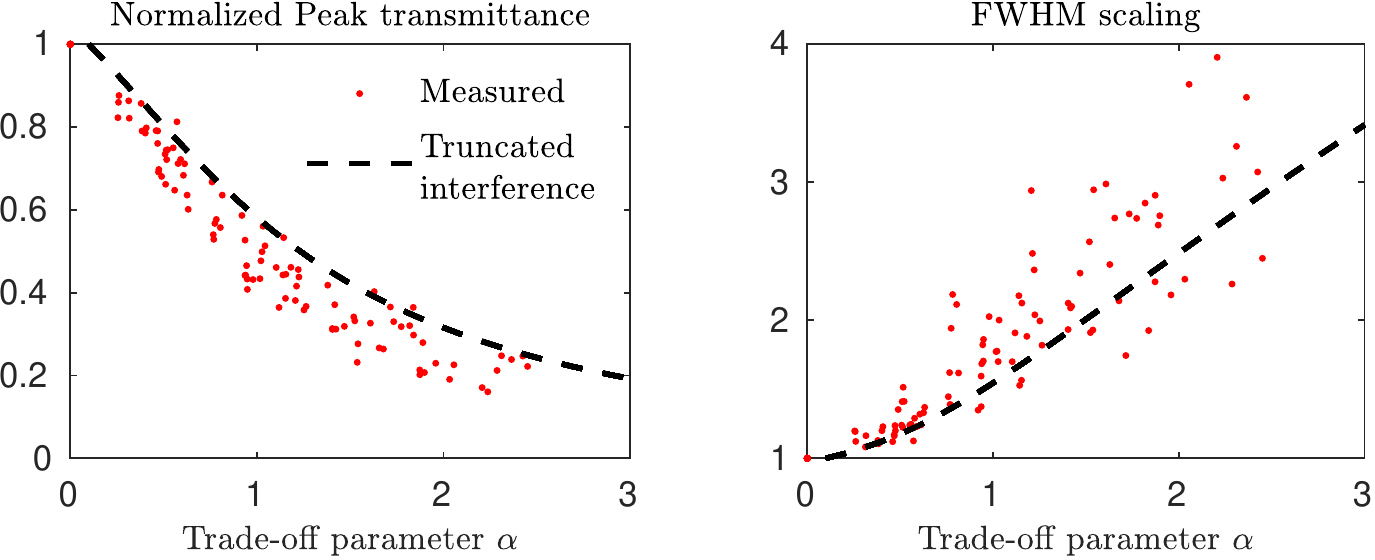}
    \caption{\label{fig:allbands} Trade-off visualized for all 25 filters on imec's snapshot mosaic sensor.}
\end{figure}

The measured changes in peak transmittance and FWHM follow the expected curve well (Fig.~\ref{fig:allbands}).
The point cloud for FWHM is wider partly because the filters are not always nicely Lorentzian due to, for example, cross-talk.

\section{Discussion} 
\label{sec:discussion}

\subsection{Mitigation strategies for snapshot spectral imaging}

\eqref{eq:tradeoff} outlines a scaling law for pixel-integrated Fabry-Pérot which constrains which camera optics can be used and what filter transmittance can be designed.
However, the impact on spectral imaging always depends on the application requirements.
Most impact is expected for applications that depend on the accurate measurement of narrowband spectral features or identification of small spectral differences between multiple samples. This is because a change in filter bandwidth will have a smoothing effect which might obscure the features of interest.

Mitigating the trade-off means designing systems for which $\alpha$ is small enough to make tiny-filter effects irrelevant.
In this section this is briefly discussed as a suggestion for future investigations. In summary, the strategies correspond to minimizing $\Theta$ or maximizing $W$ and $\Lambda_\infty$ whenever possible.
 
A first way of mitigating the trade-off is to minimize the incidence angle $\theta$ by using high f-number telecentric lenses. This is however a tight constraint for applications that require fast lenses with small f-numbers (e.g. in low light conditions) or wide-angle lenses. Furthermore, even at normal incidence there is a width-dependent transmittance reduction (Fig.~\ref{fig:peaklaw}).

A second strategy is to use materials with higher refractive indices to minimize the normalized incidence angle $\Theta$. 
While this seems an obvious approach, one is often limited by available materials and processing constraints.

A third strategy is to maximize the normalized filter width $W$ by patterning the filters across larger pixel areas. Snapshot cameras with such larger filters exist but require additional lens array to replicate the scene onto each filter tile \cite{Geelen2014a}.  

A potential fourth strategy is to maximize the bandwidth $\Lambda_\infty$ and explore different filter designs. This implies using computational approaches to enable the use of broadband filters with, for example, sinusoidal transmittance \cite{Jia2015a}.
Note that the trade-off parameter was derived for narrowband filters. Whether broadband filters would indeed be more robust to tiny-filter effects remains to be explored.

Lastly, an exciting strategy to explore is computational correction. Now that a fast simulation method has been derived, it can perhaps be used to solve the inverse problem and correct spectral measurements in a cost-efficient way. Of course, it remains to be seen whether this inverse problem will have good enough conditioning.

\subsection{Model assumptions and extensions}
The proposed wave-optics model does not explicitly take into account neighboring filters and the corresponding boundary conditions. Yet, it produces good predictions of the transmittance.
It should be better understood when and how the use of the angular spectrum method breaks down.
Furthermore, all calculations were done for the absorption-free case and further analysis will be required to verify whether the presented models remain valid.

The proposed ray-optics model intuitively explains the effect of filter size and delivers remarkably accurate and fast predictions. However, the analytical expression for peak transmittance fails to predict the reduced transmittance at normal incidence because diffraction-related effects are not included. It would be valuable to generalize the peak transmittance equation to the wave-optics regime.

Both models ignore flux that arrives outside of the pixel area and which might contribute to cross-talk. Advancing the theoretical understanding of cross-talk in filter arrays is suggested for future investigations and could lead to new filter patterning strategies.

It was shown experimentally that for the tested pixel size, a 2D wave-optics model is effective in estimating the transmittance for 3D filters up to an unknown scalar. To make the analysis maximally relevant for \textit{imaging}, a three-dimensional generalization is required that models diffraction in both dimensions and can include focused light.

\subsection{Efficient toolbox}

Although Fabry-Pérot filters were the main focus in this manuscript, the wave-optics model is agnostic about the filter design. Therefore, the time- and memory-efficient wave-optics method could be a useful tool for fast pixel-integrated filter design exploration.

The software implementation is publicly available as a MATLAB toolbox \cite{tinythinfilm}. This will facilitate reproducibility and future experimental validation of the proposed models and trade-off law.

\section{Conclusions}  
\label{sec:conclusion} 
It was demonstrated that pixel size impacts the transmittance of pixel-integrated thin-film filters by limiting the number of reflections and by diffraction. The proposed analytical models enable fast transmittance predictions which were validated experimentally and numerically. This enables quick exploration of the effect of filter width; a feature lacking in current commercial thin-film software tools.

The need for higher spatial resolution for snapshot spectral imaging devices requires thin-film filters integrated on smaller pixels. However, further filter miniaturization will require a trade-off between filter bandwidth and filter size. 

While the identified trade-off is intrinsic to the technology, it remains open whether it will be the main bottleneck for further miniaturization or whether other constraints will dominate first.
Mitigation strategies that could be further explored include using higher refractive index materials, different filter patterning, the use of microlenses, and computational approaches.

\section{Acknowledgements}

My sincere gratitude goes to Prof. Claude Amra (Institut Fresnel) for teaching me the essentials of thin-film filter wave-optics analysis and for many entertaining discussions which were instrumental to the conception of this work. I also thank Dr. Peter B. Catrysse (Stanford University) for proofreading the manuscript and providing much constructive feedback. Also, this work would not have been possible without my fruitful collaboration with imec (Belgium) during my doctoral studies. Lastly, I would like to thank Brian Wandell and Joyce Farrell (Stanford University) for their support and for facilitating this work.

\section{Disclosures} 
The author was affiliated with imec until October 2020.

\section{Supplemental documents}
See Supplemental Document 1 for derivations and further details on the numerical simulations and experiments.

See Supplemental Code 1 for the script used for FDFD simulations.

% Bibliography
%\bibliography{library.bib}

%
%
%
%
%
%
%
%
%
%
%
%
%
%
%
%
%
%

%%%%%%%%%%%%%%%%%%%%%%%%%%%%%%%%%%%%%% APPENDIX  %%%%%%%%

\newpage 
\begin{appendices}
\begin{center}
 {\huge Supplementary Material}
\end{center}

\section{Introduction}
This document contains supplementary derivations and additional documentation on experiments en simulations.

\section{Transmittance of a 2D tiny filter using wave packets} 
\label{app:tinytransmittance}
In this section the transmittance of a pixel-integrated (tiny) filter is calculated using the angular spectrum.
The derivation in this chapter is original work but inspired on a derivation for the flux transfer of a wave-packet \cite[p.70]{amra2021electromagnetic}.

The novelty in this supplementary document is to limit the domain of integration to the size of the pixel which requires non-trivial modifications for practical computation.
We will concern ourselves only with the \textit{tangential} components of the field amplitudes as only these components contribute to flux transfer.
 
But first, for completeness of notation, the thin-film transfer-matrix is briefly introduced in Section A.

\subsection{Thin-film transfer-matrix method for plane waves}
\label{sec:tfm}
For the classical case of plane-wave illumination on an infinitely wide filter there exists the well-known \emph{transfer-matrix method} to calculate the transmittance \cite{macleod2017,amra2021electromagnetic}.
As conventional in the thin-film filter literature, we only consider the (complex-valued) electrical and magnetic field components that are parallel (tangential) to the thin-film filter surface, as only these components contribute to flux transfer across the interface \cite{macleod2017}. 

For plane waves we have that the tangential electrical field $\E$ and the magnetic field $\HH$ are related as
\begin{equation}
  \label{eq:app-he-eta}
  \HH  = \eta \E,
\end{equation}
with $\eta$ [$AV^{-1}$], the characteristic admittance of each layer \cite{macleod2017}. 
For the tangential components, by construction, the admittance depends on the polarization and is calculated as
\begin{equation}
  \label{eq:app-ch5-eta}
\eta=
\begin{cases}
\dfrac{1}{\chi_0 \mu_r}\dfrac{n}{\cos \theta_n} & \text{p-polarized}\, (\textup{TM})\\\\
\dfrac{1}{\chi_0 \mu_r}n \cos \theta_n & \text{s-polarized}\, (\textup{TE})
\end{cases},
\end{equation}
with \(\chi_0=\sqrt{\mu_0/\epsilon_0} = 2.6544 \times 10^{-3}\,\text{S}\) defined as the admittance of free space \cite{macleod2017} and $\theta_n$ the angle of refraction in the film, calculated using Snell's law. For optical frequencies $\mu_r\approx1$ \cite{macleod2017}. The polarization-dependent cosines are a direct consequence of mapping the actual field amplitudes onto the axis tangential to the material interface.

\begin{figure}[H] 
\centering
\includegraphics[width=0.8\linewidth]{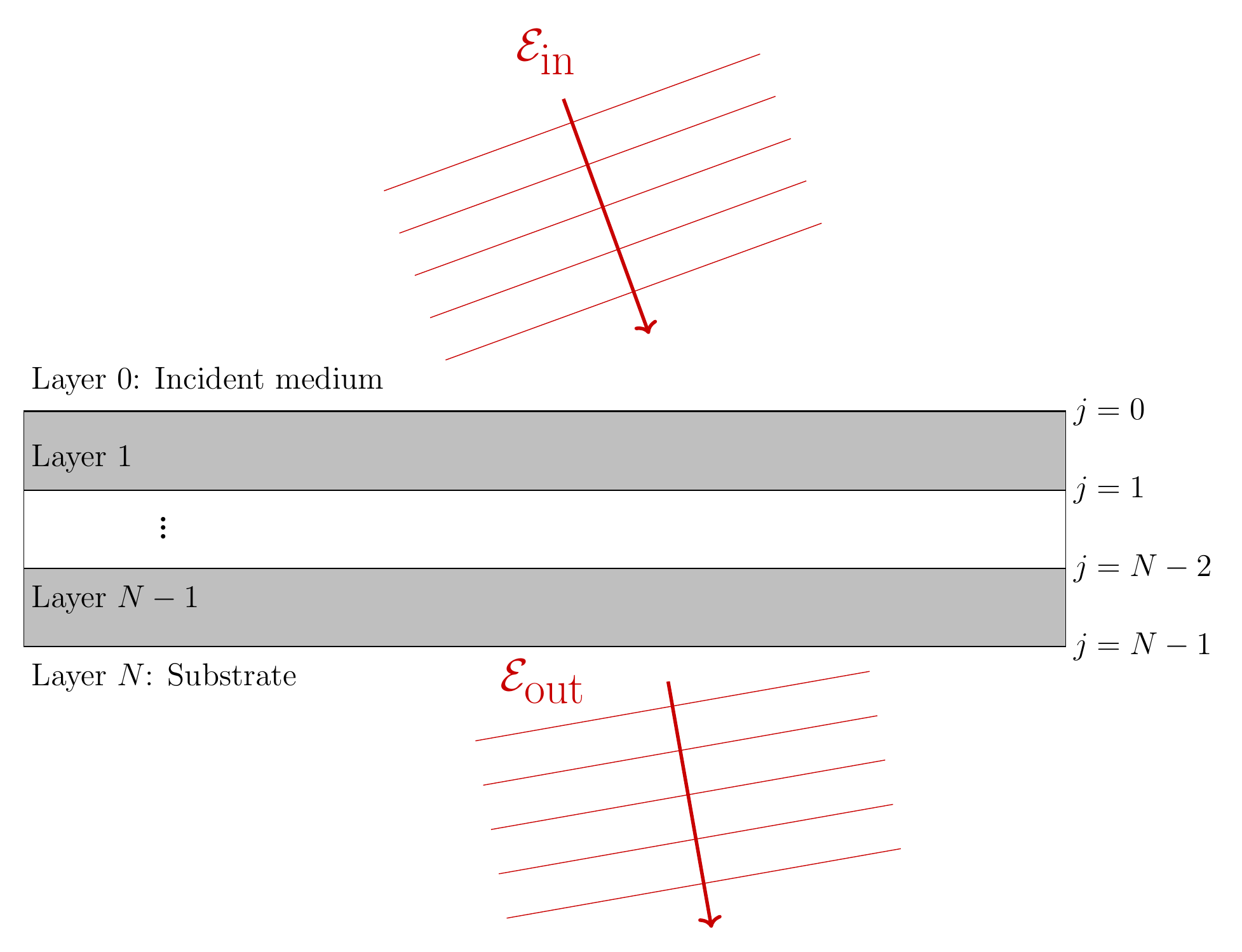}
\caption{\label{fig:inffilter} A plane wave arriving at an infinitely wide thin-film filter stack.}
\end{figure}

Consider a plane wave arriving at a stack of thin layers (Fig.~\ref{fig:inffilter}).
The plane wave will partially reflect and transmit at each material interface. Therefore, in each layer there will be forward and backward propagating waves that interfere and produce standing waves. In the substrate, there is only a forward propagating wave with amplitude $\E_{\text{out}}$.

Because the tangential components of the electrical and magnetic field remain continuous across
the material interface, each interface can be assigned a single amplitude for the 
electrical field \(\mathcal{E}_{j}\) and the magnetic field \(\mathcal{H}_{j}\). These are are recursively related by a ``transfer matrix'' so that
\begin{equation}
\label{eq:app-tfmatrix}
\begin{bmatrix}
\mathcal{E}_{j-1}\\
\mathcal{H}_{j-1}
\end{bmatrix}
=
\begin{bmatrix}
\cos \delta_j  &  \dfrac{-i}{\eta_j} \sin \delta_j \\ 
-i\eta_j\sin\delta_j & \cos\delta_j
\end{bmatrix}
\begin{bmatrix}
\mathcal{E}_{j}\\
\mathcal{H}_{j}
\end{bmatrix},
\end{equation}
with $\delta_i = \frac{2\pi n_i h_i \cos(\theta_i)}{\lambda}$ the phase thickness of each layer. 
For a stack of $N+1$ layers, this notation implies that the incident field is 
\(\mathcal{E}_\textup{in}=\mathcal{E}_0\) and the transmitted field in the
substrate is \(\mathcal{E}_\textup{t}=\mathcal{E}_{N}\) (Fig.\,\ref{fig:inffilter}).
  
One can define an equivalent admittance at each interface $j$, called the complex surface admittance \(Y_j=\mathcal{H}_j/\mathcal{E}_j\) which can be calculated using either \eqref{eq:app-tfmatrix} or the convenient recurrence relation 
\begin{equation}
Y_{j-1}=\dfrac{Y_j\cos\delta_j - i \eta_j \sin\delta_j}{\cos \delta_j - i\dfrac{Y_j}{\eta_j}\sin\delta_j},\quad\text{for}\, j=N,\hdots, 0,
\end{equation}
with \(Y_0\) being the surface admittance of the whole filter stack \cite[p.123]{amra2021electromagnetic}.
As there is no backward propagating wave in the substrate, \(Y_{N-1}=\eta_{N}\), which can be used to initiate the recursion.

In \cite[p.126]{amra2021electromagnetic} it is then derived that for a thin-film filter
stack, the transmission coefficient \emph{for the tangential components} becomes
\begin{equation}
\label{eq:app-stack-trans}
t=\dfrac{1+r}{\prod_{j=1}^{p} (\cos \delta_j - i\frac{Y_j}{\eta_j}\sin\delta_j)},\, \text{with } r=\dfrac{\eta_0 - Y_j}{\eta_0+Y_j},
\end{equation}
which is constructed so that the transmitted wave in the substrate has an amplitude \(\mathcal{E}_\text{out}=t\mathcal{E}_\text{in}\).

\subsection{Tiny filter transmittance}
Calculating the flux corresponds to calculating the irradiance and integrating it across the pixel.
For a plane wave, a simple expression exists for the irradiance such that
\begin{equation}
  \label{eq:app-Iplane}  
  I =\text{Re}(\eta)\dfrac{1}{2}\conj{\E}\E.
\end{equation}
For a wave packet, which consists of a distribution of plane waves, it is tempting to calculate the flux by summing up the irradiances for each plane wave in the decomposition. However, this approach would ignore the fact that the plane waves interfere. 

In contrast, the irradiance of an arbitrary wave is calculated as
\begin{equation}
I(x)=\dfrac{\text{Re}(\conj{\E} \mathcal{H})}{2},
\end{equation}
which is the amplitude of the Poynting vector perpendicular to the material interface.

The total flux across the tiny filter is then obtained by integrating
over the filter area:
\begin{equation}
  \label{eq:app-fluxformula}
\Phi = \int_{-w/2}^{w/2}  I(x) \dd x.
\end{equation}

Since the incoming wave is a plane wave, we can substitute \eqref{eq:app-Iplane} such that the incident flux equals
\begin{equation}
  \Phi_{\text{in}} =  \dfrac{\eta_{\text{in}}w}{2},
\end{equation}
a calculation which is not possible for the transmitted wave-packet.

To solve \eqref{eq:app-fluxformula} for a wave packet we make use of two identities.
First, for an arbitrary wave packet consisting of plane waves, we can write
\begin{equation}
\E(x) = \int_{-\infty}^\infty A(\nu) e^{i(2\pi\nu x)} \dd\nu,
\end{equation}
and
\begin{equation}
\mathcal{H}(x) = \int_{-\infty}^\infty B(\nu') e^{i(2\pi\nu' x)} \dd{\nu'}.
\end{equation}
Second, for each plane wave, the magnetic field is related by the admittance $\eta$ to the electrical
field as $B(\nu') = \eta(\nu')A(\nu')$. The admittance is polarization and angle-dependent, and hence different for each plane wave in the decomposition. Its angle-dependency is encoded by its dependency on the spatial frequency $\nu'$.

So we have that
\begin{equation}
\mathcal{H}(x) = \int_{-\infty}^\infty \eta(\nu')A(\nu') e^{i(2\pi\nu' x)} \dd{\nu'},
\end{equation}
which after substitution in \eqref{eq:app-fluxformula} gives
\begin{equation}
\Phi = \dfrac{1}{2} \int_\nu \int_{\nu'}\dd\nu \dd\nu'  \conj{A(\nu)}\eta(\nu') A(\nu') \int_{-w/2}^{w/2}\textup{d}x e^{2i\pi(\nu'-\nu)x}.
\end{equation}

To speed up numerical integration and reduce required memory, several identities are used to reduce the number of integrals.
First, we notice that the inner integral is a Fourier transform and solve it exactly so that
\begin{align}
  \label{eq:app-flux}
  \Phi =& \dfrac{1}{2} \int_\nu \int_{\nu'}\textup{d}\nu \textup{d}\nu' \conj{A(\nu)}\eta(\nu') A(\nu')  \dfrac{\sin(\pi (\nu'-\nu)w)}{\pi(\nu'-\nu)},\nonumber\\
     =& \dfrac{1}{2}  \int_{\nu'} \dd\nu' \eta(\nu')A(\nu') \int_\nu\textup{d}\nu \conj{A(\nu)} \dfrac{\sin(\pi (\nu'-\nu)w)}{\pi(\nu'-\nu)}.
\end{align}
Second, the new inner integral is in fact a convolution integral such that we can write
\begin{equation}
\Phi = \dfrac{1}{2}  \int_{\nu'} \textup{d}\nu' \eta(\nu')A(\nu') (\conj{A} * K)(\nu'),
\end{equation}
with
\begin{equation}
K(\nu')= \dfrac{\sin(\pi \nu'w)}{\pi\nu'},
\end{equation}
which I refer to as the ``pixel kernel'' in the software implementations.
When necessary, the convolution can be computed efficiently in $O(n\log n)$ time using a Fast-Fourier Transform (FFT). 

The transmitted flux becomes
\begin{align}
\Phi_{\text{t}}  =&  \dfrac{1}{2}  \int_{\nu'} \textup{d}\nu' \eta_t(\nu')A_\text{t}(\nu') (\conj{A_\text{t}} * K)(\nu').
\end{align}

We can limit the integration domain to the incident wave between $\pm 90^\circ$ incidence such that the limits of the domain are
\begin{equation}
\nu = \pm \dfrac{1}{\lambda}.
\end{equation}

Finally, we have that the transmittance of a tiny filter equals
\begin{equation}
  \label{eq:app-tinytransmittance-wave}
T(\lambda;\theta) =\dfrac{\Phi_\text{t}}{\Phi_\text{in}}.
\end{equation}

\section{Counting the number of interfering rays}
\label{app:countingM}
In the ray-optics model a converging series is truncated to model the effect of finite filter width. The index of truncation is equal to the number of interfering rays $N(x)$. This sections how this number is calculated.

\begin{figure}[H]
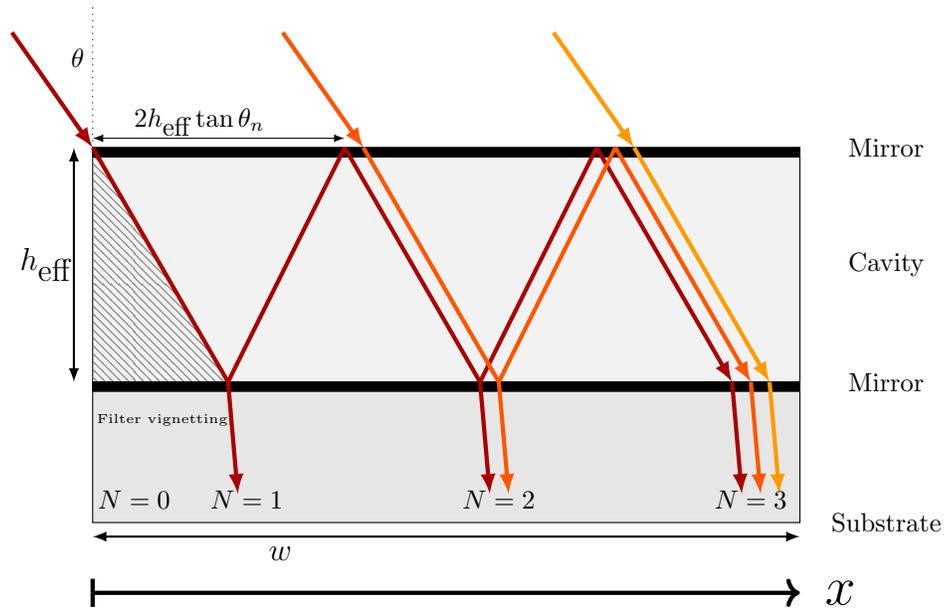

 \centering
\includestandalone[width=0.99\linewidth]{./fig/tinyfilter/reflections_tinyfilter}
\caption{\label{fig:app-tinymonolayer} Schematic illustration of the reflections in a tiny Fabry-Pérot filter on a substrate, with $N(x)$ the number of interfering rays at position $x$. The overlapping rays are drawn next to each other for illustration purposes only.}
\end{figure}

To determine the number of interfering rays $N(x)$ at position $x$, we need to analyze the number of reflections that can occur at oblique incidence. 
From Fig.\,\ref{fig:app-tinymonolayer} it can be deduced that, due to
multiple reflections, any outcoming ray must have traversed the film
\(1+2n\) times, with \(n\) a non-negative integer. 
For example, if there are two outcoming rays (\(N=2\)), at most, a ray had to traverse the film 1+2 times: the
ray is first transmitted (+1), then reflected at the bottom (+1), then reflected again at the top (+1).
The number of rays \(N(x)\) is limited by the condition that
light can reflect only a finite number of times in the film.
In other words, the total horizontal displacement of a reflected ray cannot exceed the filter dimensions, i.e., 
\begin{equation}
  x - \underbrace{d\cdot\overbrace{[1+2(N(x)-1)]}^\text{Nb. of traversals}}_{\text{total displacement}} \geq 0,
\end{equation}
with \(d=\heff\tan\theta_n\) being the horizontal displacement of a single traversal.

Solving to the largest integer value of \(N(x)\) gives 
\begin{equation}
N(x) = \left\lfloor\dfrac{1}{2} + \dfrac{x}{2d}\right\rfloor = \left\lfloor\dfrac{1}{2} + \dfrac{x}{2h\tan\theta_n}\right\rfloor.
\end{equation}
which for a half-wave plate $\heff=\frac{\lcwl}{2\neff}$, becomes
\begin{equation}
N(x) = \left\lfloor\dfrac{1}{2} + \dfrac{x\neff}{\lcwl\tan\theta_n}\right\rfloor.
\end{equation}

\begin{figure}[H]
 \includegraphics[width=0.99\linewidth]{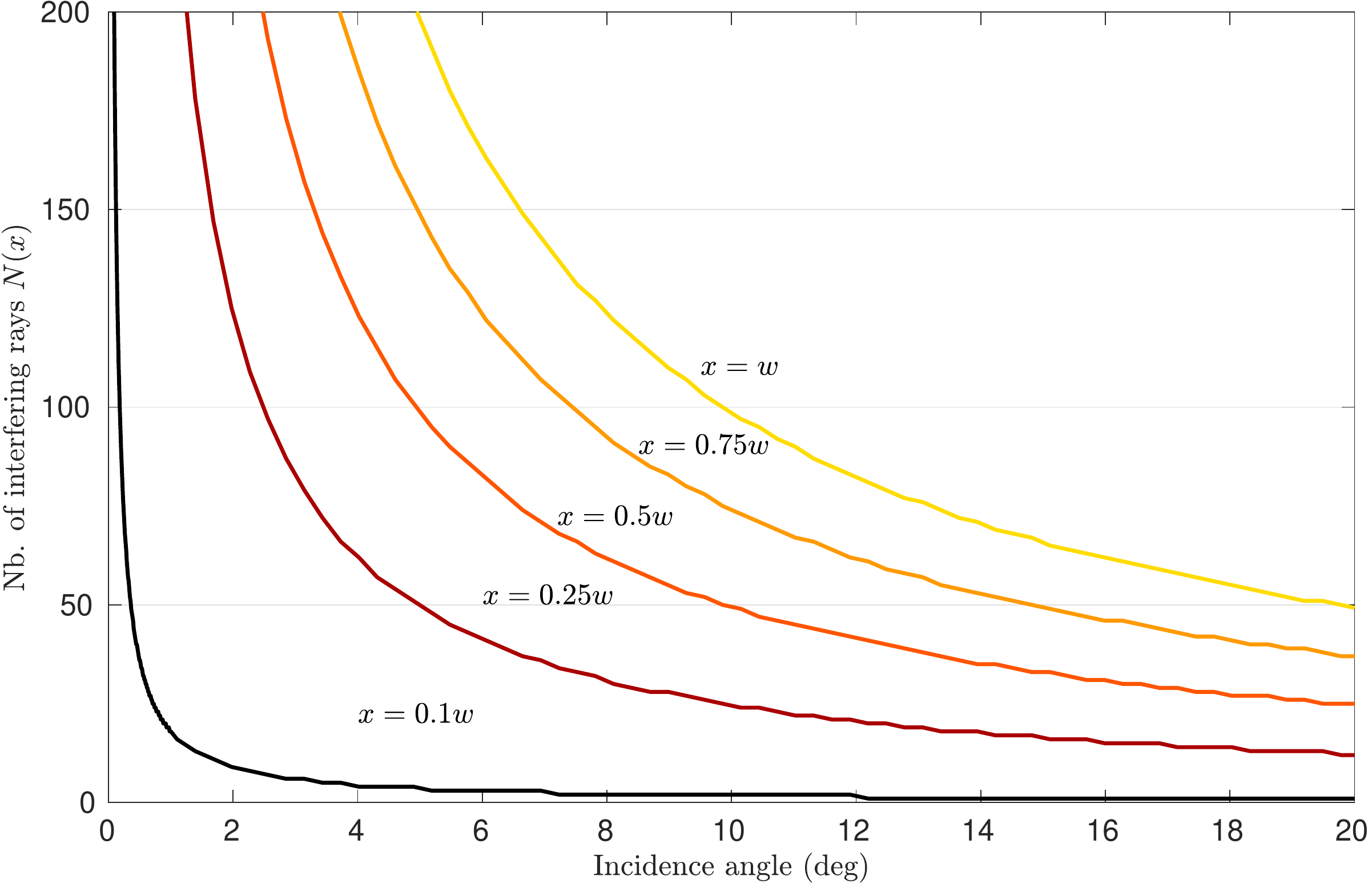}
  \caption{\label{fig:N} The spatial and angle-dependent number of interfering rays in the substrate.}
\end{figure}

The angular and spatial dependency of $N(x)$ is visualized for a filter with $\neff=1.7$ and $w=5.5$ µm in Fig.~\ref{fig:N}. It is shown for five positions on the pixel: at $x=0.1$, $x=0.25w$, $x=0.5w$, $x=0.75w$, and $x=w$.

At normal incidence, by construction, the number of interfering rays is infinite.
At oblique incidence, by contrast, the number of interfering transmitted varies across the substrate. Close to the left wall ($x=0$), the number of rays is much lower than elsewhere. Furthermore, the number of rays becomes smaller for increasing angles because the horizontal displacement \(h\tan\theta_n\) for each traversal becomes larger. This means that a reflecting ray will reach the boundary of the film with less reflections.

\section{Relationship between normalized FWHM and mirror reflectance}
\label{app:fwhm}
A key property of the ray optics monolayer model is that its response is fully determined by two dimensionless factors, the effective refractive index $\neff$ and the reflectance $R$ of the mirrors.
In this section it is shown that the mirror reflectance determines the filter bandwidth normalized by its central wavelength.
I present my own derivation because I believe it to be more transparent about the assumptions than the derivations I found in reference works like \cite[p.365]{born1999} and \cite[p.185]{macleod2017}. 

Let us consider the transmittance for an infinitely wide filter of unit peak value so that
\begin{align}
  T_{\text{peak}} = \dfrac{(1-R)^2}{R^{2}-2R \cos(2\delta) +1} .
\end{align}

The goal is to find the phase thickness for which the peak transmittance is halved (equals $1/2$).
Solving to $\delta$ we obtain
\begin{equation}
\delta_{1/2} = \dfrac{1}{2}\arccos(2-\dfrac{R}{2}-\dfrac{1}{2R}).
\end{equation}

To find the FWHM, we need to derive how a change in phase-thickness translates to a change in wavelength.
The phase thickness is non-linearly related to the wavelength as
\begin{equation}
\delta = \dfrac{2\pi \neff \heff }{\lambda} = \dfrac{\pi\lcwl}{\lambda},
\end{equation}
which can be differentiated such that
\begin{equation}
  \dd{\delta} = -\dfrac{\pi\lcwl}{\lambda^2}   \dd{\lambda}.
\end{equation}

In the neighborhood of the central wavelength $\lambda=\lcwl$ this gives
\begin{equation}
  \dd{\delta} = -\dfrac{\pi\dd{\lambda}}{\lcwl}.
\end{equation}

Assuming this equation holds for well for small enough finite ranges, we have that for $\Delta \delta = |2\delta_{1/2}|$,
\begin{equation}
  \Delta \delta  \approx \pi\dfrac{\textup{FWHM}}{\lcwl}= \pi\Lambda_\infty
\end{equation}
with \begin{equation}
  \label{eq:app-alpha}
  \Lambda_\infty = \dfrac{\text{FWHM}}{\lcwl}.
\end{equation}
the normalized bandwidth of the corresponding infinitely wide filter at normal incidence.

After substitution, we obtain the expression for the normalized filter bandwidth
\begin{subequations}
\begin{eqnarray}
  \Lambda_\infty &\approx& \dfrac{1}{\pi} \arccos(2-\dfrac{R}{2}-\dfrac{1}{2R}),\\
          &\sim& \dfrac{1-R}{\pi \sqrt{R}},\, \text{for } R\rightarrow 1,\\
          &\sim& \dfrac{1-R}{\pi},\, \text{for } R\rightarrow 1,
\end{eqnarray}
\end{subequations}
which can also be solved to $R$ such that
\begin{equation}
\begin{array}{lll}
  R &=& 2 - \cos(\pi\Lambda_\infty) - \sqrt{3 - 4 \cos(\pi\Lambda_\infty) + \cos^2(\Lambda_\infty \pi)},\\
    &\sim& 1-\pi \Lambda_\infty,\, \text{for } \Lambda_\infty \rightarrow 0.
\end{array}
\end{equation}

While this approximation is very good for narrowband filters, the analysis is simpler in the wavenumber domain because the wavenumber and phase thickness are linearly related. Hence no approximation is required to relate the wavenumber-FWHM to the mirror reflectance so that
\begin{equation} 
  \delta = \dfrac{\pi k}{k_{\text{cwl}}} \Rightarrow \Delta\delta = \pi \dfrac{\Delta_k}{k_{\text{cwl}}},
\end{equation}
and 
\begin{equation}
 \dfrac{\Delta k}{k_{\text{cwl}}} = \dfrac{1}{\pi} \arccos(2-\dfrac{R}{2}-\dfrac{1}{2R}).
\end{equation}
In this work I opted to work in the wavelength domain because it is most familiar to the target audience.

\section{Transmitted irradiance using a ray model}
  \label{app:raytransmit}
This section is an extended discussion and derivation of the ray-optics equations introduced in Section 4 of the main manuscript.
  
The correct operation of a thin-film interference filter depends on light beams reflecting many times and interfering with each other. When the filter is infinitely wide, there is an infinite number of rays that interfere.
In contrast, when the filters are tiny, the supply of rays at oblique incidence is limited but increases with distance from the wall (Fig.~\ref{fig:tinymonolayer}). This means that the number of interfering rays varies across the substrate.

This interpretation is formalized by calculating the resulting field amplitude at each position. For classical thin-film theory, this leads to a converging infinite series \cite[p.361]{born1999}. However, when there are only \(N(x)\) outcoming rays at position $x$, the series has to be truncated such that
\begin{equation}
\label{eq:app-seriestrans}
\E_t(x;\lambda) =  t_s t^2 \sum_{n=1}^{N(x)} R^{(n-1)}  e^{i2\delta(n-1)},
\end{equation}
where $t$ is the transmission coefficient of the mirror, $r$ the reflection coefficient $R=r^2$ and $t_s$ the transmission coefficient of the air-substrate interface.
The coefficients $t$ and $r$ are related as $\conj{t}t+\conj{r}r=T+R=1$

This series can be rewritten as a geometric series for which a closed
expression exists such that \cite{Zwillinger2014} 
\begin{align}
\E_t(x;\lambda) =& \left(\dfrac{t_st^2}{R e^{i\delta}}\right) \sum_{n=1}^{N(x)} (\underbrace{R e^{i2\delta}}_q)^n, \nonumber\\
	      =&  \left(\dfrac{t_st^2}{R}\right) \dfrac{(q^{N(x)}-1)}{q-1}, \label{eq:app-geomtrans}
\end{align}
which is equal to
\begin{align}
  \E_t(x;\lambda)    &= t_s t^2\cdot  \left( \dfrac{R^{N}e^{i2N\delta}-1}{Re^{i2\delta}-1}\right).
\end{align}

The transmitted irradiance is calculated as
\begin{align}
  I_t &= \dfrac{\eta_s}{2} \overline{\E_t} \E_t \\
      &=\dfrac{\eta_s}{2} \conj{\left[ t_s t^2\cdot  \left( \dfrac{R^{N(x)}e^{i2N\delta}-1}{Re^{i2\delta}-1}\right)\right]} {\left[ t_s t^2\cdot  \left( \dfrac{R^{N(x)}e^{i2N\delta}-1}{Re^{i2\delta}-1}\right)\right]}\\
      &=\dfrac{\eta_s}{2} \left[ \conj{t_s} \conj{t}^2\cdot  \left( \dfrac{R^{N(x)}e^{-i2N\delta}-1}{Re^{-i2\delta}-1}\right)\right]  {\left[ t_s t^2\cdot  \left( \dfrac{R^{N(x)}e^{i2N\delta}-1}{Re^{i2\delta}-1}\right)\right]}\\
      &= \dfrac{\eta_s}{2} \underbrace{(\conj{t}t)^2}_{T^2=(1-R)^2} \underbrace{(\conj{t_s}t_s)}_{T_s} \cdot \dfrac{(R^{N(x)}e^{-i2N\delta})(R^{N(x)}e^{i2N\delta})}{(R^{N(x)}e^{-i2\delta})(R^{N(x)}e^{i2\delta})}
  \end{align}

One can show that
\begin{align}
\left(R^{N(x)}e^{-i2N\delta}-1\right) \left(R^{N(x)}e^{i2N\delta}-1\right)= R^{2N}-2R^N \cos(2N\delta) +1,
\end{align}
from which follows that the transmitted irradiance equals
\begin{align}
  I_t(x) &= \left(\dfrac{\eta_s T_s T^2}{2}\right)\cdot \left(\dfrac{R^{2N}-2R^N \cos(2N\delta) +1}{R^{2}-2R \cos(2\delta) +1}\right)\\
      &= \dfrac{\eta_s}{2}\cdot T_s (1-R)^2 \left(\dfrac{R^{2N(x)}-2R^N(x) \cos(2N(x)\delta) +1}{R^{2}-2R \cos(2\delta) +1}\right)
\end{align}
with $T=1-R$.

\begin{figure}[H]
 \includegraphics[width=0.99\linewidth]{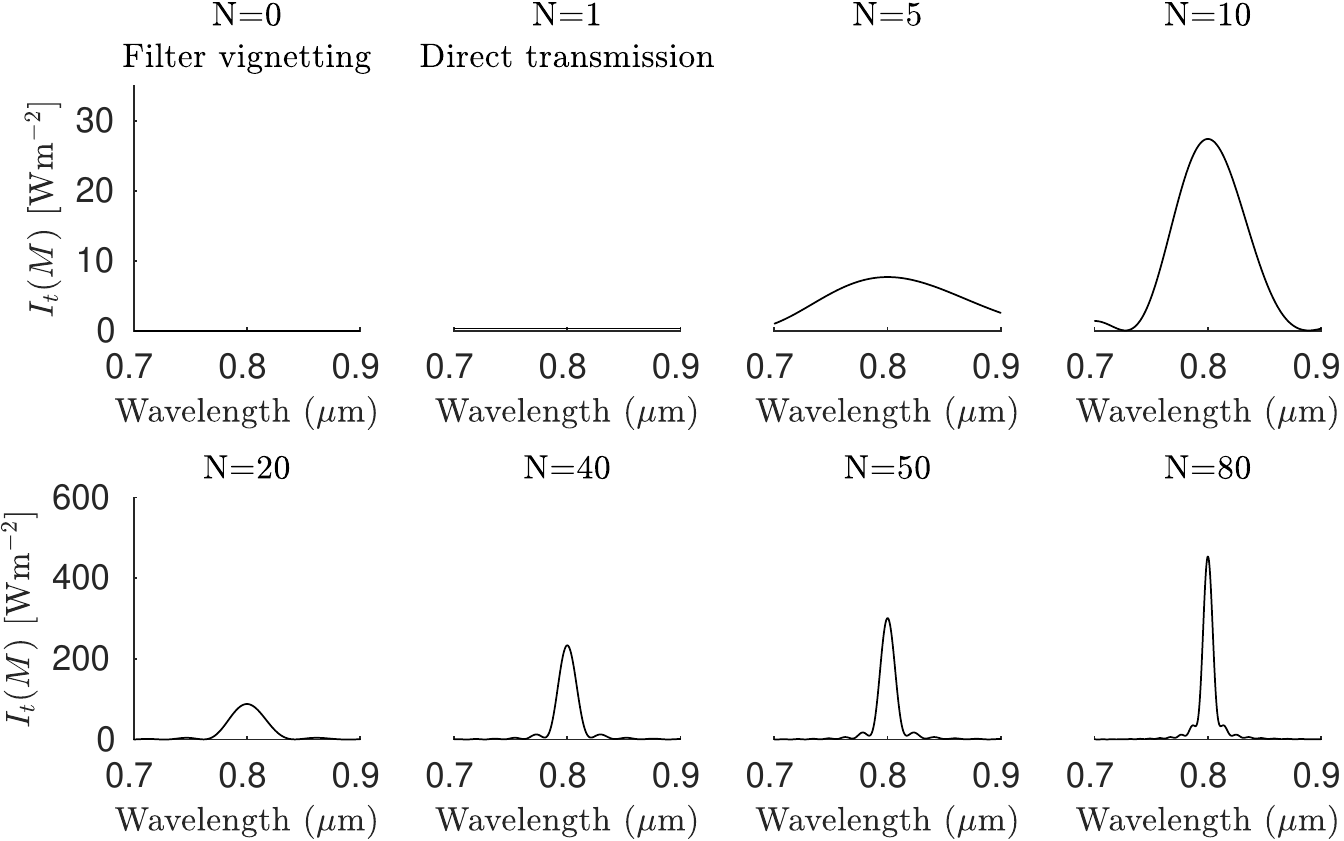}
  \caption{\label{fig:mvalues} The irradiance of a Fabry-Pérot filter for a given number of interfering rays.}
\end{figure}

The transmitted irradiance is visualized as a function of $N$ for a filter with $\neff=1.7$ and $w=5.5$ on a substrate with $n_{\text{sub}}=3.67$ (Fig.~\ref{fig:mvalues}). The more rays, the higher and more narrowband the irradiance becomes.

The total transmitted power to the pixel is obtained by integrating $I_t(x)$ across the filter. This corresponds to taking a weighted sum of irradiances visualized in Fig.~\ref{fig:mvalues}.
An analytical approximation for this is given in the next section.

\section{Analytical approximation of the transmittance}
\label{app:analytical}
In this section a very accurate approximation for the transmittance of a tiny filter is derived. This equation enables fast computation and the derivation of an expression for peak transmittance. 

Consider the transmitted irradiance in terms of the number $N(x)$ of interfering rays such that
\begin{equation}
  I_t (N(x))= \dfrac{\eta_s}{2}\cdot T_s (1-R)^2 \left(\dfrac{R^{2N}-2R^N \cos(2N\delta) +1}{R^{2}-2R \cos(2\delta) +1}\right),
\end{equation}
with.
\begin{equation}
  \label{eq:app-app-M}
  N(x) = \left\lfloor\dfrac{1}{2} + \dfrac{\neff}{\lcwl\tan\theta_n}x\right\rfloor.
\end{equation} 

The total incident flux is determined by integrating the irradiance over the pixel area so that
\begin{equation}
\Phi_t = \int_0^w I_t(x) \textup{d} x.
\end{equation}
Given that N is constant for a certain region on the pixel we can decompose the integral as
\begin{equation}
\Phi_t = \sum_{N=1}^{M-1} \int_{x_N}^{x_{N+1}} I_t(x) \textup{d} x,
\end{equation}
where $M$ is the maximum number of rays that will interfere (the furthest from the boundary), i.e. $M=N(w)$.
Consequentially, the irradiance in each region will also be a constant such that we can write
\begin{align}
  \Phi_t &= \sum_{N=1}^{M-1} \int_{x_N}^{x_{N+1}} I_t(N) \textup{d} x,\\
         &= \sum_{N=1}^{M-1} I_t(N) \int_{x_N}^{x_{N+1}}  \textup{d} x,\\
         &= \sum_{N=1}^{M-1} I_t(N) \left(x_{N+1}-x_N\right),\\
         &\approx \sum_{N=1}^{M-1} I_t(N) \Delta x,
\end{align}

For a cavity with a thickness much smaller than its width, the horizontal displacement $\Delta_x=\left(x_{n+1}-x_n\right)$ is small enough to make a continuum approximation. This is the case when $h = \dfrac{\lcwl}{2\neff} \ll w$, or equivalently, $\lcwl \ll 2\neff w$.
Under this continuum assumption we have from \eqref{eq:app-app-M},
that $\textup{d}N = \dfrac{\neff}{\lcwl\tan\theta_n} \textup{d}x$ and $\Delta N = \dfrac{\neff}{\lcwl\tan\theta_n} \Delta x$ such that
\begin{equation}
  \Phi_t \approx \dfrac{\lcwl\tan\theta_n}{\neff}  \sum_{N=1}^{M-1} I_t(N) \Delta N.
 \end{equation}

Making the continuum approximation we obtain
\begin{align}
  \Phi_t \approx \dfrac{\lcwl\tan\theta_n}{\neff}  \int_0^M I_t(N) \textup{d} N,
\end{align}
where we neglected the contributions for $N=0$ and $N=M$ to obtain a simpler equation. This is a good approximation as long as $M$ is large.

Equivalently we can now define the incident flux as
\begin{align}
  \Phi_{\text{in}}  &\approx  \dfrac{\lcwl\tan\theta_n}{\neff}  \int_0^M I_{\text{in}}  \textup{d}N =  \dfrac{\lcwl\tan\theta_n}{\neff} I_{\text{in}} \int_0^M   \textup{d} N\\
                   &= \dfrac{\lcwl\tan\theta_n}{\neff} M I_{\text{in}} 
\end{align}
So finally we have that the transmittance is
\begin{align}
  \label{eq:app-tray}
  \Tray &=\dfrac{  \Phi_{t}  }{  \Phi_{\text{in}}  } = \dfrac{\displaystyle \int_0^M I_t(N) \textup{d} N}{M}\\
        &=  \dfrac{\eta_s}{\eta_{\text{in}}}\cdot T_s (1-R)^2 \dfrac{\left[1 + \dfrac{R^{2M}-1}{\log R^{2M}} - \dfrac{2(\log(R) (R^M\cos(2M\delta)-1)+ 2\delta R^M\sin(2M\delta))}{4M\delta^2+M\log(R)^2}\right]}{R^{2}-2R \cos(2\delta) +1}.
\end{align}

Because of the continuum approximation, the function $\Tray$ is not periodic anymore but centered around $\delta=0$. This is similar, although vice versa, to the periodicity introduced by discretizing a continuous Fourier transform to perform a discrete Fourier transform.
Therefore, when using the original definition of the phase thickness $\delta= \pi\lcwl/\lambda$, one has to subtract $\pi$ before evaluating $\Tray$.

\begin{figure}[H]
  \centering
    \begin{subfigure}[t]{0.99\linewidth}
    \includegraphics[width=0.99\linewidth]{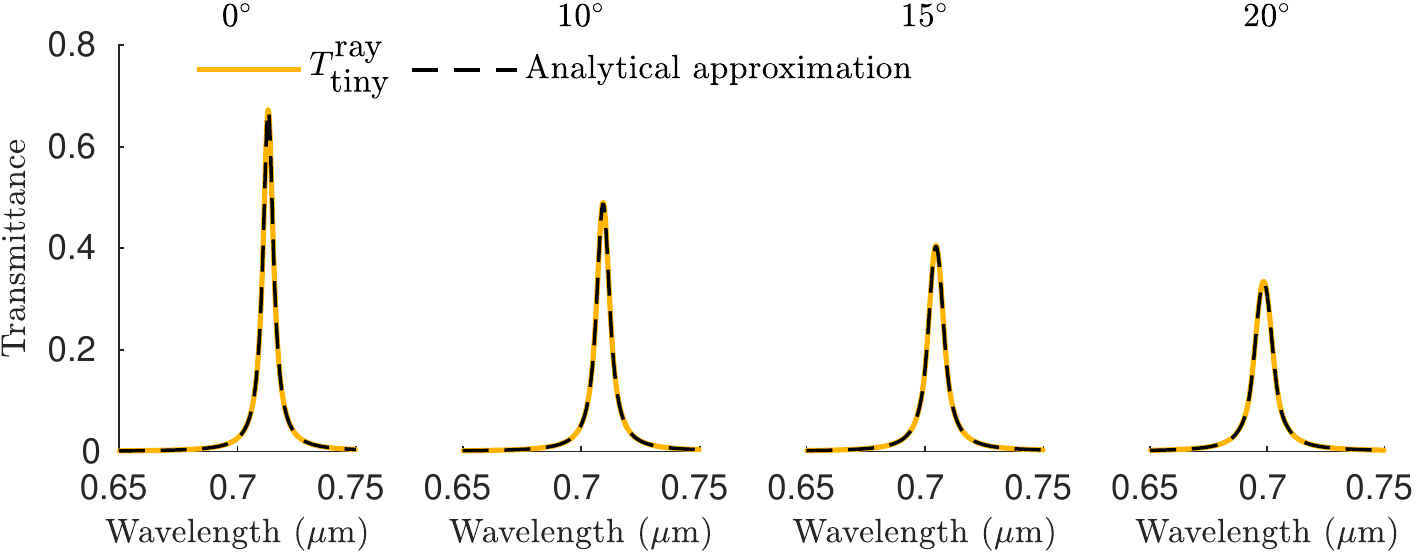}
    \caption{\label{fig:tinyray-10micron} 10 µm pixel}
  \end{subfigure}
  \begin{subfigure}[t]{0.99\linewidth} 
    \includegraphics[width=0.99\linewidth]{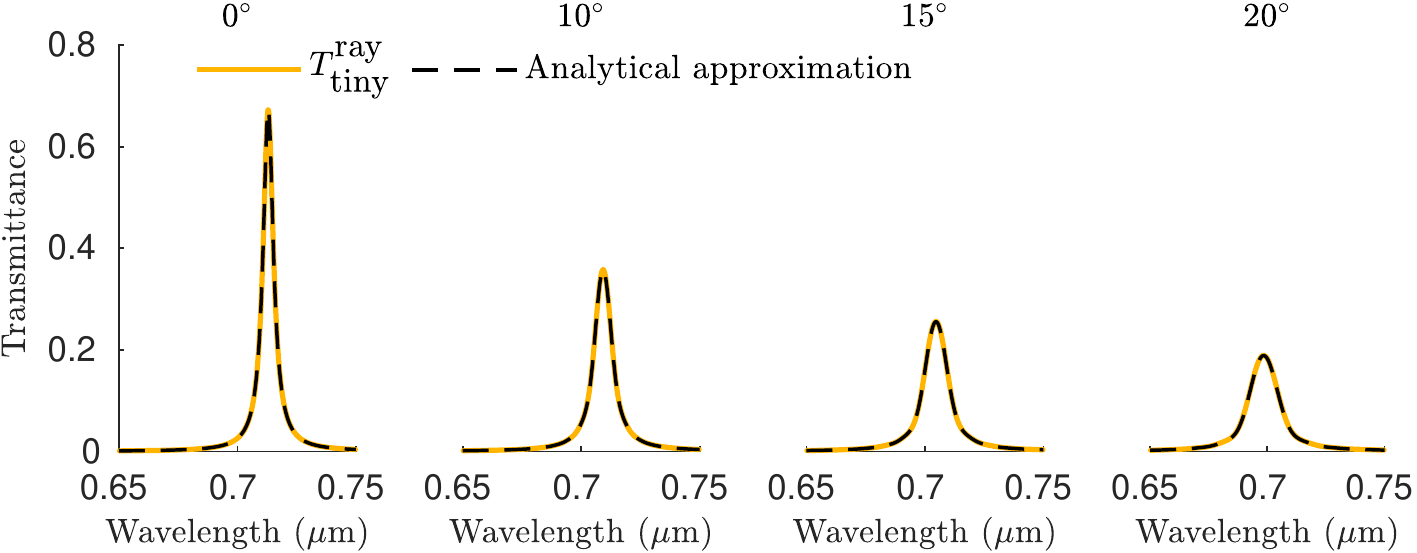}
    \caption{\label{fig:tinyray-5micron} 5.5 µm pixel}
  \end{subfigure}
  \begin{subfigure}[t]{0.99\linewidth}
    \includegraphics[width=0.99\linewidth]{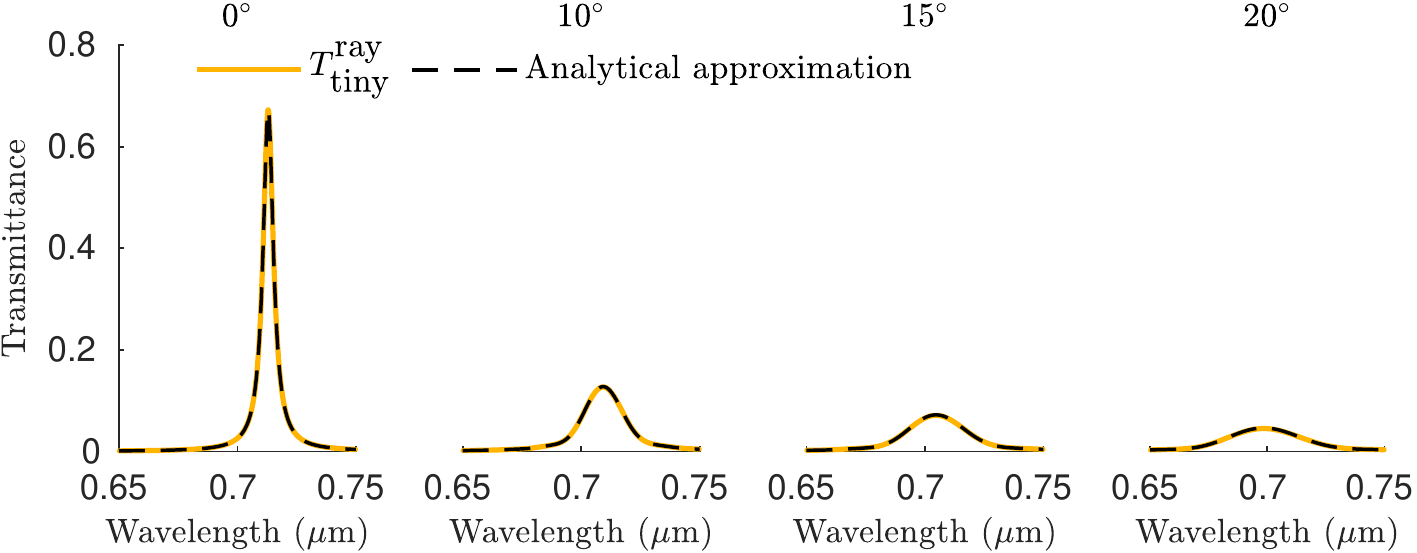}
    \caption{\label{fig:tinyray-2micron} 2 µm pixel}
  \end{subfigure}
  \caption{\label{fig:tinyray} The analytical approximation matches perfectly with the numerical prediction.}
\end{figure}

The high accuracy of this approximation is displayed in Fig.~\ref{fig:tinyray}.

\section{Peak-transmittance approximation}
\label{app:peak}
\subsection{General approximation}
To obtain the peak transmittance we evaluate \eqref{eq:app-tray} at $\delta=0$ such that
\begin{align}
  T_{\text{peak}}^{\text{ray}}   &=  \dfrac{\eta_s}{\eta_{\text{in}}}\cdot T_s \left[1+ \dfrac{(1-R^M)(3-R^M)}{\log R^{2M}}\right],
\end{align}
which can be normalized as
\begin{align}
  \dfrac{T_{\text{peak}}^{\text{ray}}}{T^\infty_{\text{peak}}}   &= 1+ \dfrac{(1-R^M)(3-R^M)}{\log R^{2M}}.
\end{align}
with $T^\infty_{\text{peak}}= T_{\text{peak}}^{\text{ray}}(M\rightarrow\infty)$. 

\subsection{Narrowband filter approximation}
To facilitate analysis for narrowband Fabry-Pérot filters we make some additional approximations.

\begin{eqnarray}
  \dfrac{T_{\text{peak}}^{\text{ray}}}{T^\infty_{\text{peak}}}  &=&1+ \dfrac{(1-e^{M\log R})(3-e^{M\log R})}{e^{2M\log R}}
\end{eqnarray}

First let us take
\begin{equation}
M \approx \dfrac{\neff w}{\lcwl\tan\theta_n} = \dfrac{\neff}{\tan \theta_n} \underbrace{\dfrac{w}{\lcwl}}_W,
\end{equation}
and
\begin{equation}
  \log R \sim \log (1+\pi\Lambda_\infty) \sim \pi\Lambda_\infty,\,\text{for} \Lambda_\infty \rightarrow 0,
\end{equation}
so that
\begin{equation}
\label{eq:app-peakvalue}
  \dfrac{T_{\text{peak}}^{\text{ray}}}{T^\infty_{\text{peak}}}   \sim 1- \dfrac{(1-e^{-\pi M\Lambda_\infty})(3-e^{-\pi M\Lambda_\infty})}{2\pi M\Lambda_\infty}.
\end{equation}

We can then rewrite the peak-transmittance formula as
\begin{equation}
   \dfrac{T_{\text{peak}}^{\text{ray}}}{T^\infty_{\text{peak}}} = 1+\dfrac{(1-e^{\pi/\alpha})(3-e^{\pi/\alpha})}{2\pi/\alpha}
 \end{equation}
 with
\begin{equation}
  \begin{array}{lcl}
  \label{eq:app-tradeoff}
  \alpha&=&\dfrac{1}{M \cdot \Lambda_\infty} = W^{-1}\cdot\Lambda_\infty^{-1} \cdot \Theta\\\\
        &=&   \underset{\substack{\text{Normalized}\\\text{spatial width}\\\\W}}{\Bigl(\underbrace{\dfrac{w}{\lcwl}}_{}\Bigr)^{-1}} \cdot
  \underset{\substack{\text{Normalized}\\\text{bandwidth}\\\\\Lambda_\infty}}{\Bigl(\underbrace{\dfrac{\text{FWHM}}{\lcwl}}_{}\Bigr)^{-1}}
  \cdot
  \underset{\substack{\text{Normalized}\\\text{incidence angle}\\\\\Theta}}{\Bigl(\underbrace{\dfrac{\tan \theta_n}{\neff}}_{}\Bigr)}
\end{array}.
\end{equation}

\section{Additional details for the trade-off law verification}
\label{app:tradeoff}
The four filter designs used to generate the curves in Figs.~5 and 6 are Fabry-Pérot filters of the form
  \begin{equation*}
  \text{Air}| H (LH)^b | (LL) | (HL)^b H|\text{Si},
\end{equation*}
with
\begin{equation*} 
  \begin{array}{lllll}
    \Lambda_\infty=2\%:    &b=4,&n_h=1.88,& n_l=1.3,    &\neff =1.47,\\
    \Lambda_\infty=1\%:    &b=5,&n_h=1.99,& n_l=1.4,    &\neff =1.57,\\
    \Lambda_\infty=0.5\%:  &b=8,&n_h=2.02,& n_l=1.58,   &\neff =1.73,\\
    \Lambda_\infty=0.25\%: &b=20,&n_h=2.09,& n_l=1.9,   &\neff=1.98.\\
  \end{array}
\end{equation*}
Some of these designs are not practically useful but used for illustration purposes only.

\section{Additional details for the numerical FDFD simulation}
\label{sec:FDFD}
The wave-optics method is validated by solving Maxwell's equations using a finite-difference frequency-domain (FDFD) Matlab Toolbox \cite{maxwellfdfd}. The frequency domain toolbox, used with a direct solver, is used for two reasons. First, we are only interested in the harmonic regime and hence time steps are not required. Second, the toolbox is very convenient for define rectangular structures like thin-film filters.

\begin{figure}[H]  
  \includegraphics[width=0.99\linewidth]{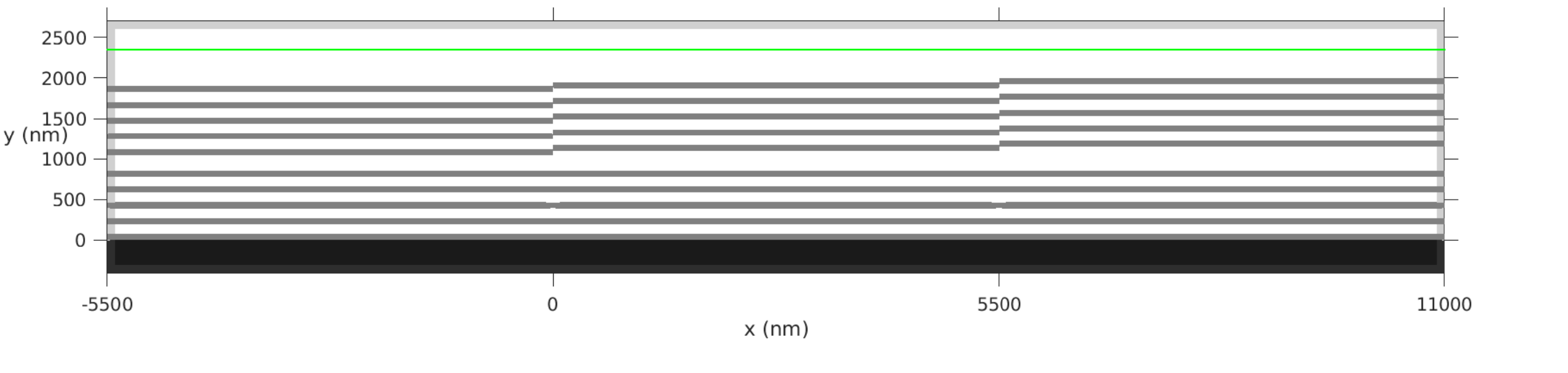}
  \caption{\label{fig:layout} FDFD simulation domain with three adjacent Fabry-Pérot filters on a single substrate.}
\end{figure}

Three different Fabry-Pérot filters are placed adjacently as in Fig.~\ref{fig:layout}.
The three designs are
\begin{equation}
  \begin{array}{lc}
    \text{Left}:&\text{Air}|H (LH)^4 L_1 L (HL)^4 H|\text{Substrate}\\
    \text{Central}:&\text{Air}|H (LH)^4 L L (HL)^4 H|\text{Substrate}\\
    \text{Right}:&\text{Air}|H (LH)^4 L_3 L (HL)^4 H|\text{Substrate}
  \end{array}
\end{equation}
Where $L$ and $H$ are quarter-wave layers for a central wavelength around $\lcwl=720$ nm. By construction, this means that the dielectric mirror has a central wavelength around 720 nm.
The material parameters that were used are $n_\text{air}=1$, $n_l=1.5$, $n_h=2.4$, and $n_\text{sub}=3.67$.

The central wavelength of the filter can be chosen by varying the cavity thickness. 
By choosing layers $L_1$ and $L_3$ to be 50 nm thinner and thicker than $L$ respectively we obtain 
central wavelengths for the left, central, and right filter of 664 nm, 720 nm, and 776 nm respectively. These were chosen to be far enough such that there is no spectral overlap and cross-talk can be, supposedly, ignored.

The grid size is 6 nm and the perfectly matched layers (PML) are 100 nm on each border. In addition, an s-polarized plane wave source is placed above the filters. The simulation domain is visualized in Fig.~\ref{fig:layout}. 

To ensure realistic boundary conditions only the the transmittance of the central filter is calculated.
The incident flux is calculated in the absence of any filter and substrate.
The transmitted power is calculated by integrating the the power flux across the width of the filter.
Depth-dependent absorption in the pixel is not modeled as this work focuses on calculating how much light is transmitted to the pixel, not modeling its quantum efficiency.

The script for this simulation is available as Supplemental Code 1.
 
% Visualization of absolute value
% \begin{figure}[H]
% \includegraphics[width=0.99\linewidth]{/home/thomas/Documents/tinyfilters/articles/tinymultilayer/fig/simulations/FDFD/fig/kleur-20deg.png}
%   \caption{\label{fig:kleurtjes} FDFD simulation.}
% \end{figure}

\section{Additional details about the experimental setup}
\label{app:experiment}
 
\subsection{Setup}

The experimental setup consists of a monochromator coupled to a collimating lens which illuminates the sensor. The setup and its operation is described in \cite{Agrawal2016}. Several components of the setup were updated but the operational principle remains unchanged. The image sensor board is placed onto a goniometer to control the incidence angle of the collimated light.

For each angle, this setup measures the spectral response of each pixel in digital numbers.
The device under test is imec's snapshot 5x5 mosaic camera \cite{Geelen2014}. On this sensor, 25 filters are integrated on 5.5 µm wide pixels of a CMV2000 image sensor \cite{CMOSIS}. The filters are distributed across a wavelength range of 665 and 975 nm.

%\begin{figure}[H]
%\centering
%  \begin{subfigure}[t]{0.49\linewidth}
%  \includegraphics[width=0.99\linewidth]{./fig/calibratie/calibratie.png%}
%  \caption{Picture of a similar setup.}
%\end{subfigure}
%\begin{subfigure}[t]{0.49\linewidth}
%  \includegraphics[width=0.99\linewidth]{./fig/calibratie/tiltstage.png}
%  \caption{Extension with goniometer.}
%\end{subfigure}
%  \caption{\label{fig:setup} Experimental setup with a monochromator coupled by a fiber bundle to a collimator which illuminates the sensor which is mounted on a goniometer.}
%\end{figure}
 
\subsection{Fit at normal incidence}
To apply the equivalent monolayer model to the measured data it was necessary to estimate the normalized bandwidth $\Lambda_\infty$. This is the bandwidth of the filter at normal incidence, \textit{if it were infinitely wide}, which is not known in this case.
However, using the wave-optics model one can sweep across multiple bandwidths and use the tiny filter transmittance that best matches the measurement.

\begin{figure}[H]
  \includegraphics[width=0.99\linewidth]{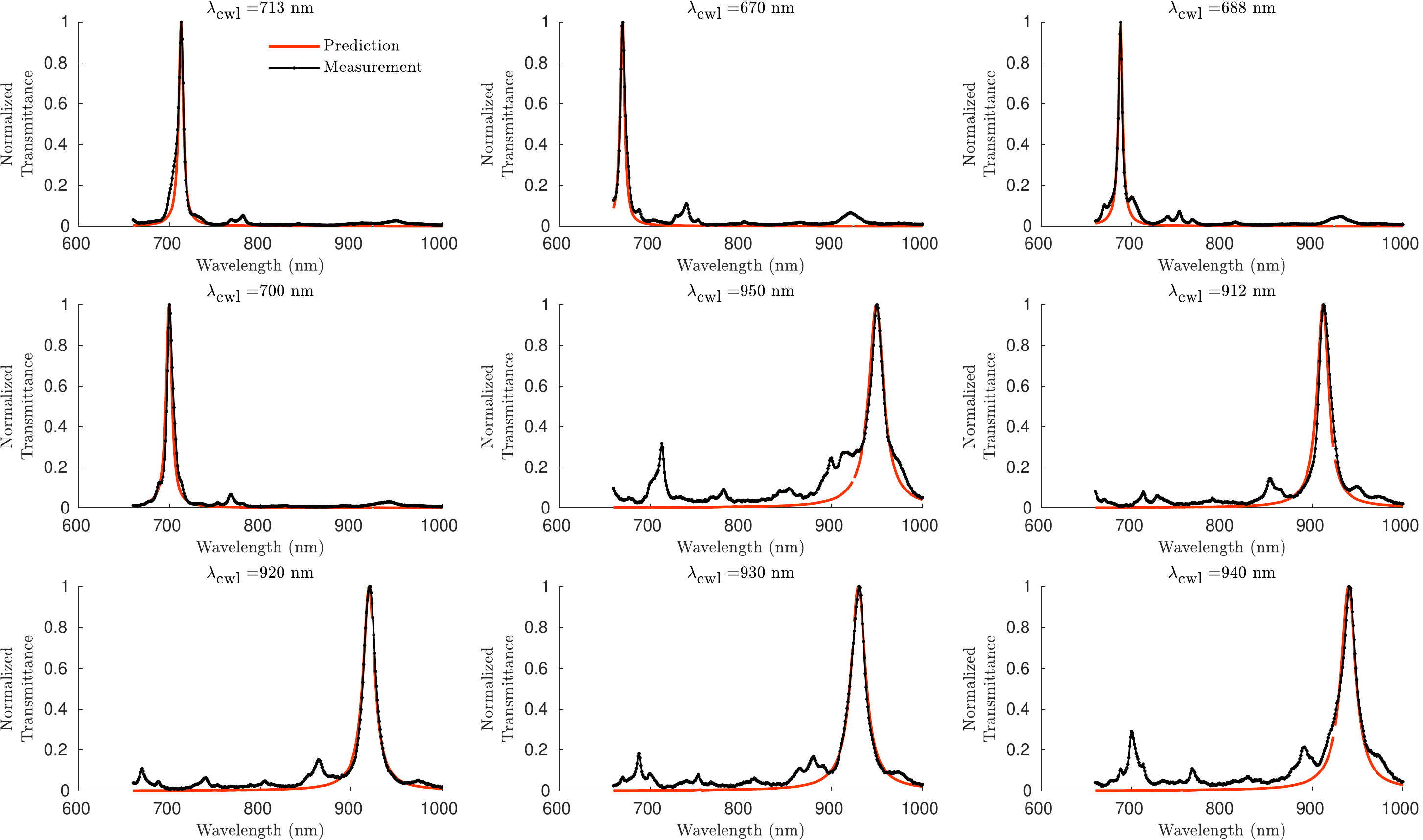}
  \caption{\label{fig:fitzero} Fit at normal incidence for several bands on imec's sensor.}
\end{figure}

The result of this procedure is displayed for several filters in Fig.~\ref{fig:fitzero}. The fit is not perfect for each filter because not each filter has a perfect Lorentzian shape. 

\subsection{Uniformity}

Because each filter is sensitive to different wavelengths, a uniformity map at a single wavelength is not practical.
Therefore, the maximal digital number (collected over all wavelengths) is used for each pixel. A 2D median 10 by 10 pixels filter is applied to smooth out the map.

\begin{figure}[H]
  \includegraphics[width=0.99\linewidth]{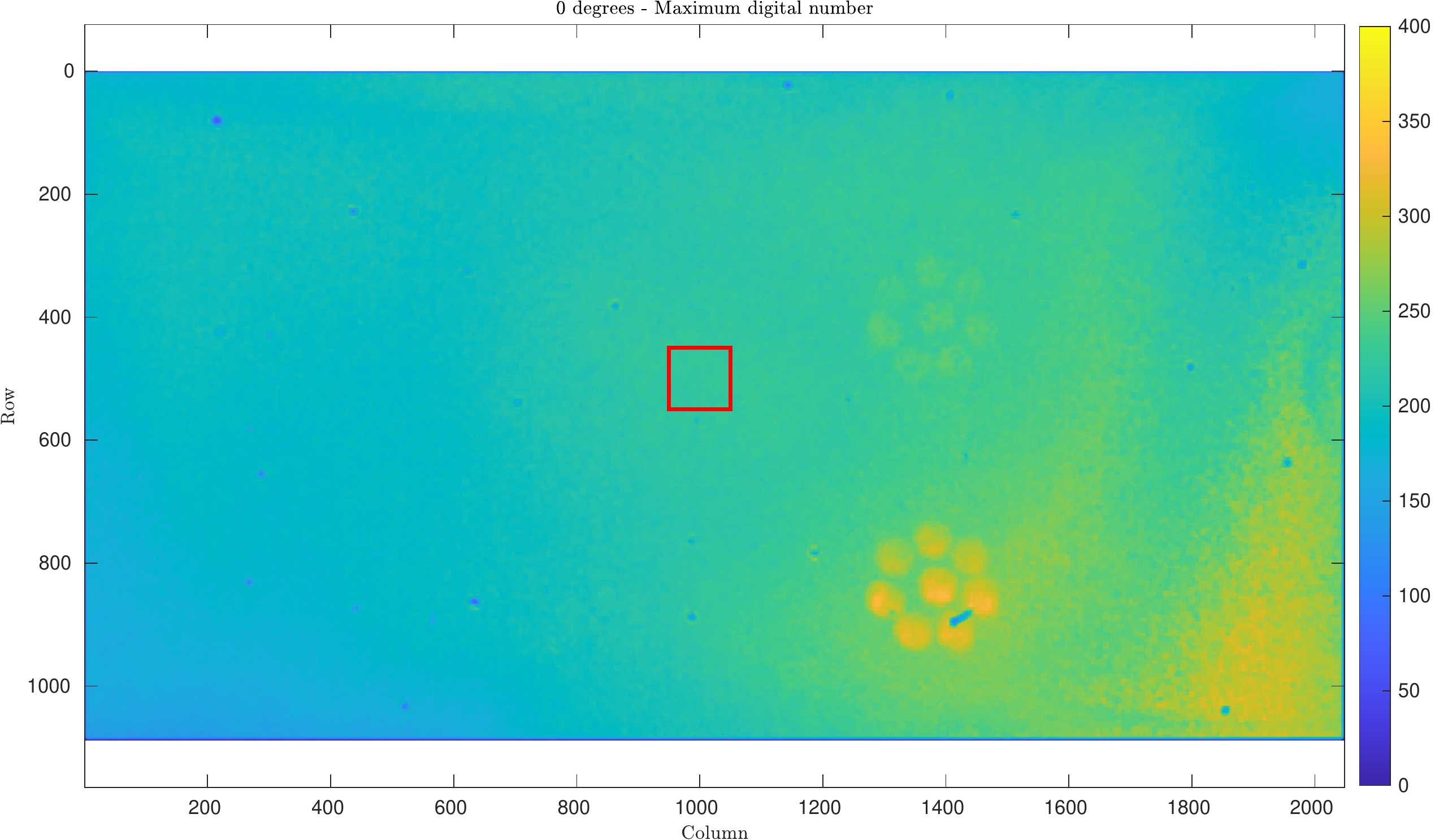}
  \caption{\label{fig:uniformity} Uniformity map. The red square was used for generating the experimental data}.
\end{figure}

At normal incidence there was a specular reflection between the sensor and the collimator which caused the multiple ``flowers'' which are an image of the optical fiber bundle (Fig.~\ref{fig:uniformity}).
Care was taken to keep this specular reflection outside the sampling region marked by the red rectangle.
At larger angles the specular reflection were not in the scene.

\begin{figure}[H]
  \includegraphics[width=0.99\linewidth]{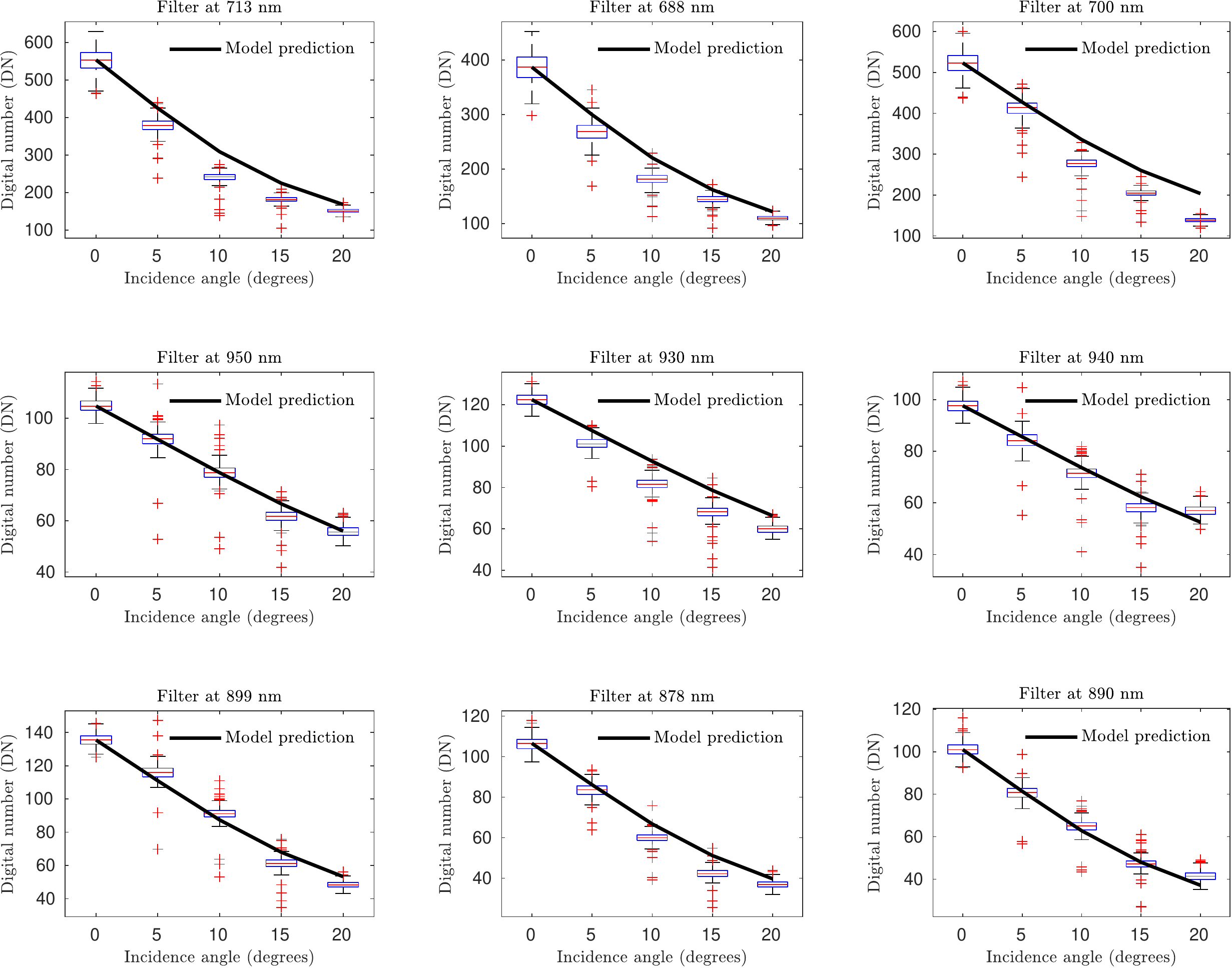}
  \caption{\label{fig:boxplots} Model predictions and uncertainty in measured region.Variations are possibly due to cross-talk or uniformity variations.} 
\end{figure}

The good fit of the (ray) model predictions to the measurements indicates that, by using the median value in the sampling region, non-uniformity was mitigated to a much smaller magnitude compared to the drop in peak transmittance for tiny filters (Fig.~\ref{fig:boxplots}). Future investigations on cross-talk will require more control over uniformity to decouple the effects.


\end{appendices}

\end{document}